\documentclass[preprints,article,accept,moreauthors,pdftex]{Definitions/mdpi} 
\firstpage{1} 
\pubvolume{1}
\issuenum{1}
\articlenumber{0}
\pubyear{2022}
\copyrightyear{2022}
\datereceived{} 
\dateaccepted{} 
\datepublished{} 
\hreflink{https://doi.org/} 

\usepackage{caption}
\usepackage{subcaption}

\usepackage{amsmath,amssymb}

\DeclareFontFamily{U}{rcjhbltx}{}
\DeclareFontShape{U}{rcjhbltx}{m}{n}{<->rcjhbltx}{}
\DeclareSymbolFont{hebrewletters}{U}{rcjhbltx}{m}{n}

\let\aleph\relax\let\beth\relax
\let\gimel\relax\let\daleth\relax

\DeclareMathSymbol{\aleph}{\mathord}{hebrewletters}{39}
\DeclareMathSymbol{\beth}{\mathord}{hebrewletters}{98}
\DeclareMathSymbol{\gimel}{\mathord}{hebrewletters}{103}
\DeclareMathSymbol{\daleth}{\mathord}{hebrewletters}{100}
\DeclareMathSymbol{\lamed}{\mathord}{hebrewletters}{108}
\DeclareMathSymbol{\mem}{\mathord}{hebrewletters}{109}
\DeclareMathSymbol{\ayin}{\mathord}{hebrewletters}{96}
\DeclareMathSymbol{\tsadi}{\mathord}{hebrewletters}{118}
\DeclareMathSymbol{\qof}{\mathord}{hebrewletters}{113}
\DeclareMathSymbol{\shin}{\mathord}{hebrewletters}{152}
\DeclareMathSymbol{\samekh}{\mathord}{hebrewletters}{115}

\Title{Temporal, structural, and functional heterogeneities extend criticality and antifragility in random Boolean networks}

\TitleCitation{Heterogeneity extends criticality and antifragility}

\Author{Amahury Jafet L\'opez-D\'iaz$^{1}$\orcidA{}, Fernanda S\'anchez-Puig$^{1,2}$\orcidB{} and Carlos Gershenson $^{2,3,4,5}$*\orcidC{}}

\AuthorNames{Amahury L\'opez-D\'iaz, Fernanda S\'anchez-Puig and Carlos Gershenson}

\AuthorCitation{L\'opez-D\'iaz, A. J.; S\'anchez-Puig, F.; Gershenson, C.}

\address{%
$^{1}$ \quad Facultad de Ciencias, Universidad Nacional Auton\'{o}ma de M\'{e}xico, 04510 Ciudad de M\'exico, Mexico\\
$^{2}$ \quad Centro de Ciencias de la Complejidad, Universidad Nacional Auton\'{o}ma de M\'{e}xico, 04510 Ciudad de M\'exico, Mexico \\
$^{3}$ \quad  Instituto de Investigaciones en Matem\'aticas Aplicadas y en Sistemas, Universidad Nacional Auton\'{o}ma de M\'{e}xico, 04510 Ciudad de M\'exico, Mexico\\
$^{4}$ \quad  Lakeside Labs GmbH, Lakeside Park B04, 9020 Klagenfurt am W\"orthersee, Austria\\
$^{5}$ \quad  Santa Fe Institute. 1399 Hyde Park Rd., Santa Fe, NM 87501, USA}

\corres{Correspondence: cgg@unam.mx}

\abstract{Most models of complex systems have been homogeneous, i.e., all elements have the same properties (spatial, temporal, structural, functional). However, most natural systems are heterogeneous: few elements are more relevant, larger, stronger, or faster than others. 
In homogeneous systems, criticality --- a balance between change and stability, order and chaos --- is usually found for a very narrow region in the parameter space, close to a phase transition.
Using random Boolean networks --- a general model of discrete dynamical systems --- we show that heterogeneity --- in time, structure, and function --- can broaden additively the parameter region where criticality is found. 
Moreover, parameter regions where antifragility is found are also increased with heterogeneity. However, maximum antifragility is found for particular parameters in homogeneous networks. Our work suggests that the ``optimal'' balance between homogeneity and heterogeneity is non-trivial, context-dependent, and in some cases, dynamic.}


\begin{document}







\section{Introduction}
Phase transitions are found in a wide range of phenomena~\cite{stanley1987introduction}, from brain dynamics~\cite{haimovici2013brain} to jazz ensembles~\cite{setzler2018creative}. Usually, critical dynamics are found near phase transitions~\cite{christensen2005complexity,Mora2011,Roli2018}. Criticality exhibits an equilibrium between order and chaos, a balance connecting robustness, information storage, variability, computation, evolvability, and adaptability. Criticality is found in a variety of complex systems and models. Antifragility is a more recent concept~\cite{Taleb2012} and can be described as a property of systems that benefit from noise, perturbations, or disorder. Antifragility has been applied in risk analysis~\cite{aven2015concept}, molecular biology~\cite{danchin2011antifragility}, and computer science~\cite{abid2014toward}, among others~\cite{jones2014engineering}.

Random Boolean networks (RBNs) were originally proposed as models of genetic regulatory networks~\cite{Kauffman1969,AldanaEtAl2003,Gershenson2004c}. They are useful when specific topologies are unknown. Moreover, since RBNs are general models, these archetypes offer us the possibility to explore the space of feasible living and computational systems. In classical RBNs (CRBNs) connections and functions are chosen randomly, but all nodes usually have the same number of connections and the same bias in their functions.  CRBNs can exhibit three dynamical regimes: ordered, critical, and chaotic. The criticality and antifragility of RBNs depend on many different factors~\cite{LuqueSole1997}. These can be exploited to guide the self-organization and evolvability of RBNs towards the critical regime~\cite{Gershenson:2010}. 

Cellular automata are particular cases of Boolean networks where the state of a variable is determined by its spatial neighbors~\cite{vonNeumann1966,Wolfram1983,WuenscheLesser1992}. When modeling ferromagnets with this kind of models it is possible to find an indication that heterogeneity modifies the dynamics. An example of this are Griffiths phases~\cite{Griffiths1969}, which arise when a certain degree of spatial disorder is applied. Under this condition, the critical temperature, $T_c$, at which finite-size scaling is observed, becomes a wide range of $T$~\cite{bray1987nature}. Griffiths phases and other rare region effects have been studied in more general models, such as complex networks~\cite{AlbertBarabasi2002,Newman:2003,NewmanEtAl2006}. Quenched disorder can be replaced by topological heterogeneity in networks with finite topological dimension~\cite{munoz2010griffiths}. These results have a wide range of implications for propagation phenomena and other dynamical processes in networks. Much of network science~\cite{Barabasi2016} studies the effect of non-trivial (i.e., heterogeneous) topologies in different properties of complex systems.

Heterogeneity is pervasive in many phenomena, from public transportation systems~\cite{gershenson2015slower, carreon2017improving} to social and biological networks~\cite{santos2006evolutionary, zhou2020universal}. Twenty years ago, scientists started exploring alternative topologies in RBNs~\cite{OosawaSavageau2002}. In 2003 Aldana showed that RBNs with a scale-free topology expand the advantages of the critical regime into the ordered phase~\cite{Aldana2003}, i.e., a scale-free topology expands the range of the critical regime. Later, Gershenson and colleagues studied the role of redundancy in the robustness of RBNs. They concluded that redundancy of nodes (a particular type of heterogeneity) prevents mutations from propagating in RBNs~\cite{GershensonEtAl2006}. Heterogeneity is common, but how relevant is it for the dynamics of complex systems? It seems that more than what was previously thought~\cite{Sanchez-Puig2022}, also highlighting the importance of nonlinearity. This principle (heterogeneity extends criticality) seems to apply to a broad range of situations. 

One example of structural heterogeneity is modularity~\cite{Schlosser:2004,callebaut05}, as it provides a level of organization that promotes at the same time robustness and evolvability. Since changes do not propagate as easily between modules in the case of RBNs, modules broaden the range of the critical regime towards the chaotic phase~\cite{BalpoGershenson:2010}. Another example of the importance of structural heterogeneity is found in the evolution of biological networks. By modifying the physical interactions in them and the way in which constraints are imposed on the states of their nodes, interesting dynamic changes are observed~\cite{smith2015potential}. It is also possible to take advantage of heterogeneity  in human-built structures. For instance, power grid networks can produce heavy tailed distributions without a self-organizing mechanism~\cite{odor2018heterogeneity}. This demeanor is due to the fact that power grids are weighted, hierarchical modular networks and particular heterogeneous topologies can be exploited to avoid global failures.

Spatial disorder generates Griffiths phases which, in analogy to critical points, are characterized by a slow relaxation of the order parameter and divergences of quantities such as the susceptibility. However, for dimensions greater than or equal to $2$, it is possible to induce Griffiths phases through time heterogeneity using absorbing states~\cite{vazquez2011temporal}. Temporal Griffiths phases are characterized by generic power-law scaling of some magnitudes and generic divergences of the susceptibility. Temporal heterogeneity is also prevalent in rank dynamics~\cite{Iniguez2021}, where few more relevant elements change slower than most less relevant elements in a variety of systems, possibly facilitating systems at the same time with robustness and adaptability.

The above is a non-exhaustive list of relevant examples of heterogeneity. Some years ago, Moretti and Muñoz showed the existence of Griffiths phases in synthetic hierarchical networks and also in empirical brain networks such as the human connectome~\cite{moretti2013griffiths}. According to them, stretched critical regions produce higher functionality in a generic way, facilitating the task of self-organizing, adaptive and evolutionary mechanisms that select criticality. Recently, Ratnayake and colleagues pointed out that complex networks with heterogeneous nodes have a greater robustness as compared to networks with homogeneous nodes~\cite{math9212769}. Also, Sormunen, et al. showed adaptive networks with critical manifolds that can be navigated as changes are made in parameters~\cite{Sormunen2022}, i.e., criticality can be associated to a manifold in a multidimensional system, instead of being restricted to a single value.

\section{Methods}
\subsection{Random Boolean Networks}
Although Boolean networks were first proposed in 1969 by Stuart A. Kauffman~\cite{Kauffman1969}, they became popular models only decades later~\cite{drossel2008random}. Boolean networks are a generalization of cellular automata. Although one might think that they are a crude simplification of genetic regulatory networks, it turns out that there are several cases where this general model does, in fact, correctly describe expressed and suppressed genetic patterns.

\begin{figure}[ht] 
        \centering \includegraphics[scale=0.2]{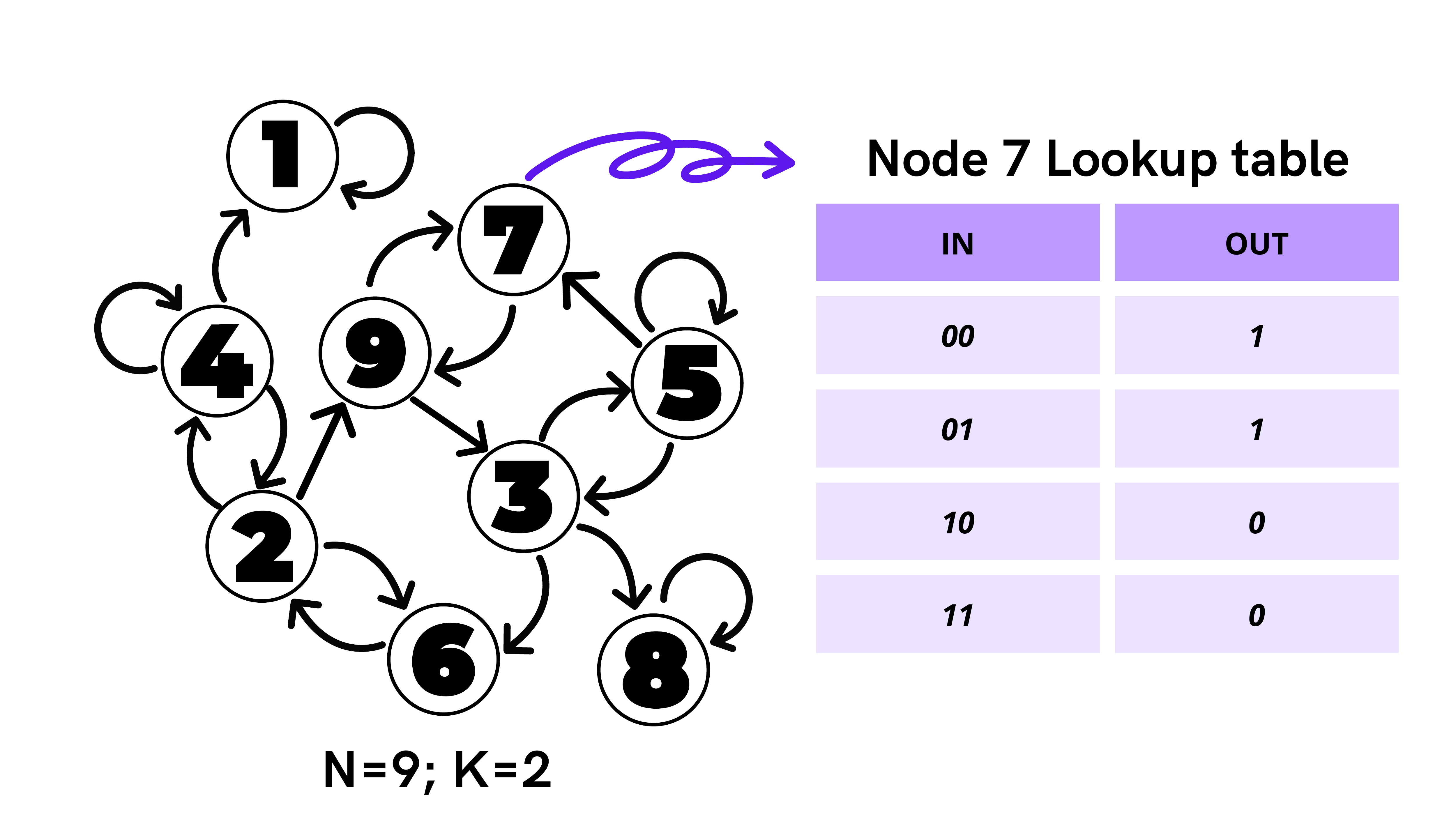}
        \caption{Illustration of a random Boolean network with $N = 9$ nodes and $K = 2$ inputs per node (self-connections are allowed). The node rules are commonly represented by lookup-tables, which associate a $1$-bit output (the node's future state) to each $2^K$ possible $K$-bit input configuration. The out-column is commonly called the ``rule'' of the node.}
        \label{RBNExample}
\end{figure}

A directed graph $D$ consists of a set $V$ of elements $a, b, c, \dots$ called the nodes of $D$ and a set $A$ of ordered pairs of nodes $(a, b)$, $(b, c)$, $\dots$ called the arcs of $D$. We use the symbol $ab$ to represent the arc $(a, b)$. If $ab$ is in the arc set $A$ of $D$, then we say that $a$ is an in-neighbour of $b$, and also that $b$ is an out-neighbour of $a$. We say that $D$ is $k$-in regular $(k \geq 1)$ if every node has exactly $k$ in-neighbours. The out-degree of a node $a$ is the number of out-neighbours of $a$. Let $D = (V, A)$ be a $k$-in directed graph, a family $\left(f_a\right)_{a\in V}$ of functions $f_a:\{0,1\}^k\longrightarrow\{0,1\}$ is called a Boolean network on $D$. Figure~\ref{RBNExample} is an example of a Boolean network on a $2-in$ directed graph with $9$ nodes. 

A Boolean network is called random if the assignment $a\mapsto f_a$ is made at random by sampling independently and uniformly from the set of all the $2^{2^K}$ Boolean functions with $K$ inputs. A function $\phi: V\longrightarrow\{0,1\}$, $a\mapsto \phi_a$, is called a state of the random Boolean network on $D$. The value $\phi_a$ is called the state of $a$. The updating function $F(\phi)$ of a state $\phi$ is the function $F:V\longrightarrow\{0,1\}$, $a\mapsto \phi'_a$, defined as
\begin{equation}
    \phi'_a = f_a\left(\phi_{a_1},\dots,\phi_{a_k}\right)
\end{equation}

Given the complexity exhibited by real genetic networks, Kauffman assumed the following four conditions in his model: 
\begin{enumerate}
\item Each gene is regulated by exactly $K$ other genes.
\item The $K$ genes that regulate each node are chosen randomly using uniform probability.
\item All genes are updated synchronously, i.e., at the same timescale \cite{Gershenson2004b}.
\item Each gene is expressed (i.e., its Boolean value is $1$) with probability $p$ and is unexpressed with probability $1-p$.
\end{enumerate}

\begin{figure}[ht] 
        \centering \includegraphics[scale=0.4]{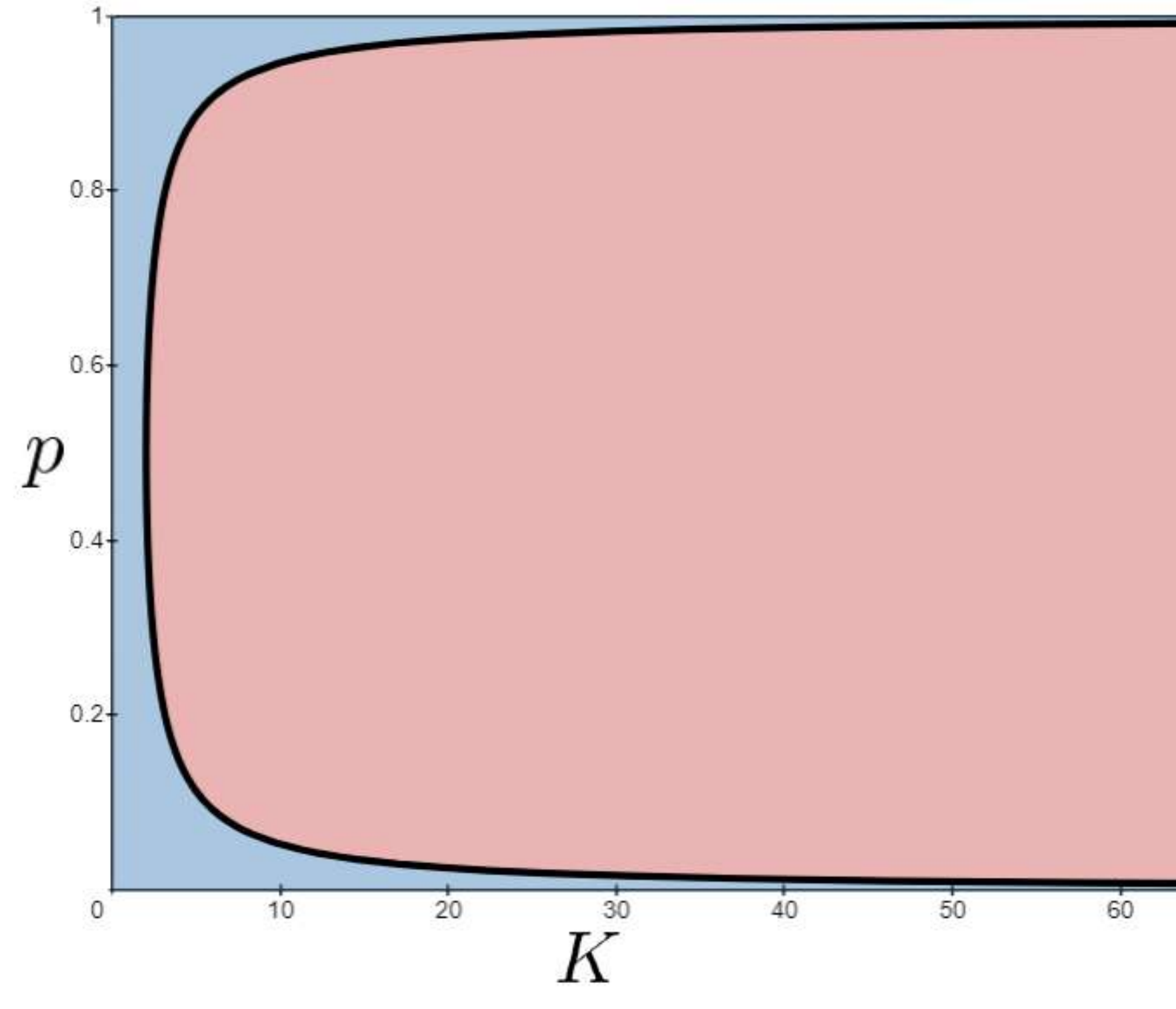}
        \caption{Phase diagram described by Derrida and Pomeau~\cite{DerridaPomeau1986}. The solid curve represents the critical connectivity $K_c = [2p(1-p)]^{-1}$. When $K<K_c$ the network exhibits ordered dynamics (small shaded area to the left of $K_c$), while for $K>K_c$ the network exhibits chaotic dynamics (large shaded area to the right of $K_c$).}
        \label{KvsP}
\end{figure}

The first two imply \emph{structural} homogeneity, the third \emph{temporal} homogeneity, and the fourth \emph{functional} homogeneity. Although these conditions might seem to limit the model, a number of interesting dynamic behaviors have been found. In particular, Derrida and Pomeau showed analytically that there exists a dynamical phase transition controlled by the parameters $K$ and $p$~\cite{DerridaPomeau1986}, assuming the thermodynamic limit, i.e., $N \rightarrow \infty$. In fact, for each value of $p$ there exists a critical connectivity $K_c=[2p(p-1)]^{-1}$ such that all perturbations in the initial state vanish if $K<[2p(p-1)]^{-1}$ and a small perturbation in the initial state propagates throughout the system for $K>[2p(p-1)]^{-1}$. The first case is referred to as the ordered phase and the second case as the chaotic phase. Figure~\ref{KvsP} shows such a phase transition for the classical version of random Boolean networks.

\begin{figure}[ht] 
        \centering \includegraphics[scale=0.6]{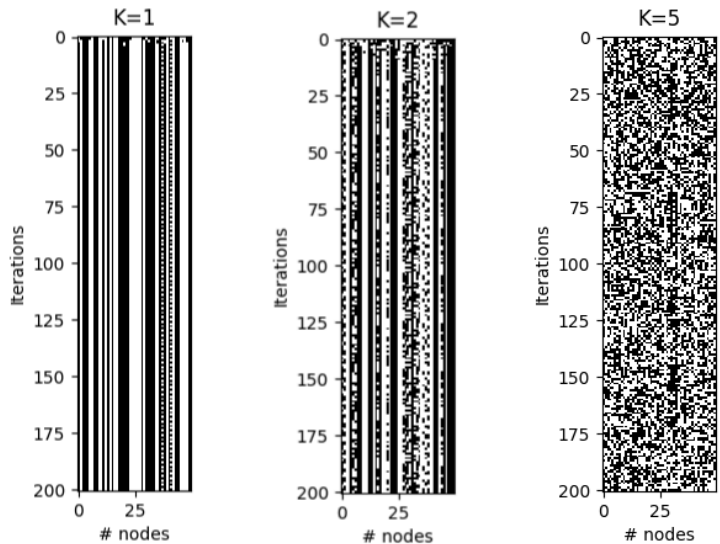}
        \caption{Example of three regimes of CRBN using 50 nodes ($N=50$) with 200 steps each. (time flows downwards).
        The resulting complexities $C$ (see Eq.~\ref{eq:C}) are: for $K=1$, $C=0.0513$,
        for $K=2$, $C=0.8651$, and
        for $K=5$, $C=0.3079$.}
        \label{crbn}
\end{figure}

Therefore, depending on different parameters, the dynamics of RBNs can be classified as ordered, critical (near the phase transition), and chaotic. Figure~\ref{crbn} shows examples of these dynamics for different $K$ values. It is important to emphasize that the model described so far is homogeneous. As can be seen in Figure~\ref{KvsP}, if criticality is only found near phase transitions, then most of the parameter space will have ``undesirable'' dynamics. To extend criticality (and also antifragility) in RBNs, we will add different types of heterogeneity to the system, which are described below.

A generalized Boolean network on a directed graph D consists of a family $(f_a)_{a\in V}$ of functions $f_a:\{0,1\}^{k_a^{-}}\longrightarrow\{0,1\}$ with $k_a^{-}\geq 1$ the in-degree of $a$. Thus, a heterogeneous random Boolean network is a generalized Boolean network whose connectivities are chosen using a particular probability distribution. In order to obtain a RBN with heterogeneous structure, the methodology proposed in~\cite{Sanchez-Puig2022} will be followed and the Poisson and exponential distributions will be used. Strictly speaking, both distributions are heterogeneous. However, the exponential is “more heterogeneous” than the Poisson. Therefore, we consider the former as the heterogeneous structural case and the latter as the homogeneous structural case.

According to the classification proposed by Gershenson~\cite{Gershenson2002e}, classical RBNs have a synchronous (homogeneous) temporality, while Deterministic Generalized Asynchronous RBNs have an asynchronous (and therefore heterogeneous) nature. The heterogeneous updating function of a state $\phi$ of a random heterogeneous Boolean network on $D$ is the function $\Tilde{F}(\phi):V\times\mathbb{N}\longrightarrow\{0,1\}$, defined by
\begin{equation*}
    \left(a,t\right) \mapsto 
\begin{cases}
\phi'_a &\text{if $t$ is a multiple of $K_a^{+}$}\\
\phi_a &\text{otherwise}
\end{cases}
\end{equation*}
where $t$ is called the discrete time-step, and $K_a^{+}$ is the out-degree of $a$. Thus, each node is updated every number of time steps equal to its out-degree, so the more nodes one node affects, the slower it will be updated. Until now, structural and temporal heterogeneity have been introduced; the impact of these two on the dynamics of RBNs has already been studied in other works. To conclude this subsection we will present a new type of heterogeneity, which we will call functional heterogeneity. 

As mentioned above, one of the conditions that Kauffman considered when setting up his model was that each gene is expressed with probability $p$ (and therefore, is unexpressed with probability $1-p$). This value of $p$ is the same for all nodes, and is fixed in the sense that it does not change throughout the dynamics; in fact, it can be shown that if $p=0.5$, then $K_c=2$. It is at this value that critical dynamics are observed (near the transition between chaos and order). Functional heterogeneity consists of assigning a probability $p_a$ to each node in the network. For this, we take the values from some probability distribution whose mean is a fixed value, in this work $0.5$. Thus, each gene is expressed with a different likelihood and as we will see later, this enriches the dynamics.

\subsection{Criticality}
Complexity, like life or consciousness, has no fully accepted definition~\cite{lloyd2001measures}. Probably, if one asks 100 complexity scientists for the definition of complexity, they will receive 150 different answers. There is a broad variety of notions and measures of complexity proposed under different contexts~\cite{Edmonds:1999}. Regardless of the area of study, a partial definition of complexity is expected to be able to capture the transfer of information between components and relate it directly to other abstractions such as emergence or self-organization~\cite{ProkopenkoEtAl2007,ComplexityExplained}. 

Inspired by~\cite{LopezRuiz:1995}, in this paper we will make use of a complexity measure based on Shannon's entropy~\cite{Shannon1948}.  Let $b$ be the length of the alphabet (for all the cases considered in this paper, $b = 2)$ and $K = 1/\log_{2}b$ a normalization factor, the Shannon entropy is given by the following expression:
\begin{equation}
    I = -K\sum_{i=1}^{b} p_i\log{p_i}. 
    \label{eq:I}
\end{equation}
Actually, chaotic dynamics are characterized by a high $I$, while ordered (static) dynamics are characterized by a low $I$. Following~\cite{Fernandez2013Information-Mea}, since the critical dynamics is characterized by having a balance between order and chaos, the complexity will be calculated through the product between $I$ and its complement,
\begin{equation}
    C = 4I(1-I),
    \label{eq:C}
\end{equation}
where the constant $4$ is added to normalize the measure to $[0, 1]$~\cite{10.3389/frobt.2017.00010}. More precisely, the complexity of each node is calculated from its temporal series, and then the complexities of all nodes are simply averaged to calculate the complexity of the network. This measure used is maximized at phase transitions and its value decreases gradually, so it is easier to calculate compared to other complexity measures, such as Fisher's information. To qualitatively estimate the amount of criticality in a random Boolean network, we will adopt the methodology presented in~\cite{Sanchez-Puig2022} and construct the complexity curves with respect to average connectivity $K$. Thus, if the maximum expected complexity is shifted, then the criticality is extended in the sense that the area under the curve is broadened. In other words, critical-like dynamics are found for a broader range of $K$ values. As shown in~\cite{Sanchez-Puig2022}, when structural or temporal heterogeneity are added to a network, then criticality increases. 

\subsection{Antifragility}
According to Taleb, antifragility is the ability to improve the capacity of a system in the face of external disturbances~\cite{Taleb2012}. Antifragility should not be confused with resilience or robustness, as these two allude to the system maintaining its functionality or properties after perturbations, whereas antifragility necessarily implies a benefit of the system in the face of disturbances. In Greek mythology, we find antifragility in the Lernaean Hydra, whose virtue of regenerating two heads for every one that was amputated reflects the idea of antifragility. In the natural world the canonical example is found in the immune system, whose ability to strengthen itself when exposed to pathogens can be seen as antifragile.

Although the concept of antifragility has been applied in many areas of knowledge, few antifragility measures have been developed to date. Following~\cite{Pineda2019}, we define the antifragility of a RBN as the product between network ``satisfaction'' gain $\Delta\samekh$ and network perturbation $\Delta x$. To define the degree of perturbation in a network of $N$ nodes we first select $X$ nodes randomly and aggregate the perturbations at a time $t$ such that $t \mod O = 0$. Thus, if $T$ is the total simulation time, the degree of perturbations is defined as follows:
\begin{equation}
    \Delta x = \frac{X\times\left(T/O\right)}{N\times T}, 
    \label{perturbation}
\end{equation}
where $0\leq \Delta x \leq 1$.

To define the level of satisfaction in a RBN we can interpret each node in the network as an agent. Since the higher the complexity, the better the balance between adaptability and robustness, we say that a node in the network is ``satisfied'' when it reaches a high level of complexity. Using the complexity measure described above (Eq.~\ref{eq:C}), the degree of satisfaction in the network is quantified as:
\begin{equation}
    \Delta\samekh = C - C_0,
    \label{satisfaction}
\end{equation}
where $C_0$ is the complexity of the network before adding disturbances and $C$ is the complexity of the network after adding disturbances. Since the value of the complexity ranges between $0$ and $1$, it follows that $-1\leq \Delta\samekh \leq 1$. Using (\ref{perturbation}) and (\ref{satisfaction}) we define the \emph{fragility} of the network as 
\begin{equation*}
    \oint = -\Delta\samekh\times\Delta x.
\end{equation*}
Since $\Delta\samekh\in\left[-1,1\right]$ and $\Delta x\in\left[0,1\right]$, then $\oint\in\left[-1,1\right]$. Note that if the satisfaction decreases with perturbations then the system is fragile, i.e. positive values of $\oint$ indicate fragility. On the other hand, if satisfaction remains constant with perturbations we say that the system is robust, from which it follows that values close to zero of $\oint$ indicate robustness. Similarly, if satisfaction increases with perturbations we say that the system is antifragile, therefore negative values of $\oint$ indicate antifragility. Since $\oint$ depends on $\Delta\samekh$, then $\oint$ can have different values depending on the state of the nodes. Thus, using multiple final states we compute the average antifragility of a RBN.




\section{Results and Discussion}
To have a simple notation we are using 'Ho' for Homogeneous, 'He' for Heterogeneous, 'S' for Structural, 'T' for Temporal and 'F' for Functional.  First, we will compare the complexity curves of the eight cases that we described above and then we are going to study the variations of that outcome using networks with different numbers of nodes. Then, we will visualize the transitions under certain conditions for each case. We will also vary some functional and one temporal parameters to study which combination of them extends criticality even further. These include the standard deviation, the mean and the domain of the functional distribution, and also we will vary the temporal update. Finally, we will explore what is the effect of different heterogeneities on antifragility.
\subsection{Homogeneity vs Heterogeneity}
\begin{figure}[htbp] 
  \begin{subfigure}[b]{0.5\linewidth}
    \centering
    \includegraphics[scale=0.5]{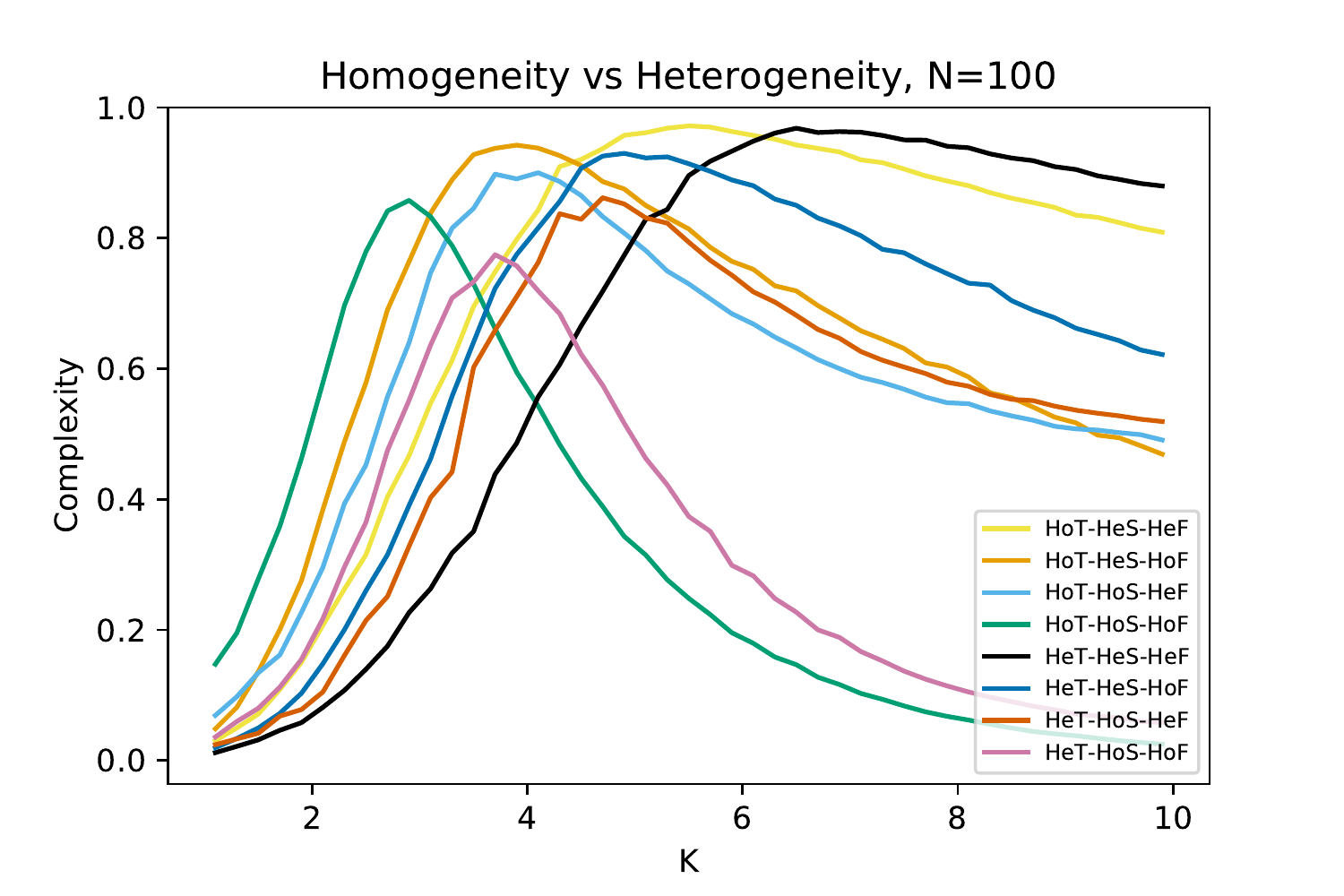} 
    \caption{} 
    \label{N100} 
    \vspace{4ex}
  \end{subfigure}
  \begin{subfigure}[b]{0.5\linewidth}
    \centering
    \includegraphics[scale=0.5]{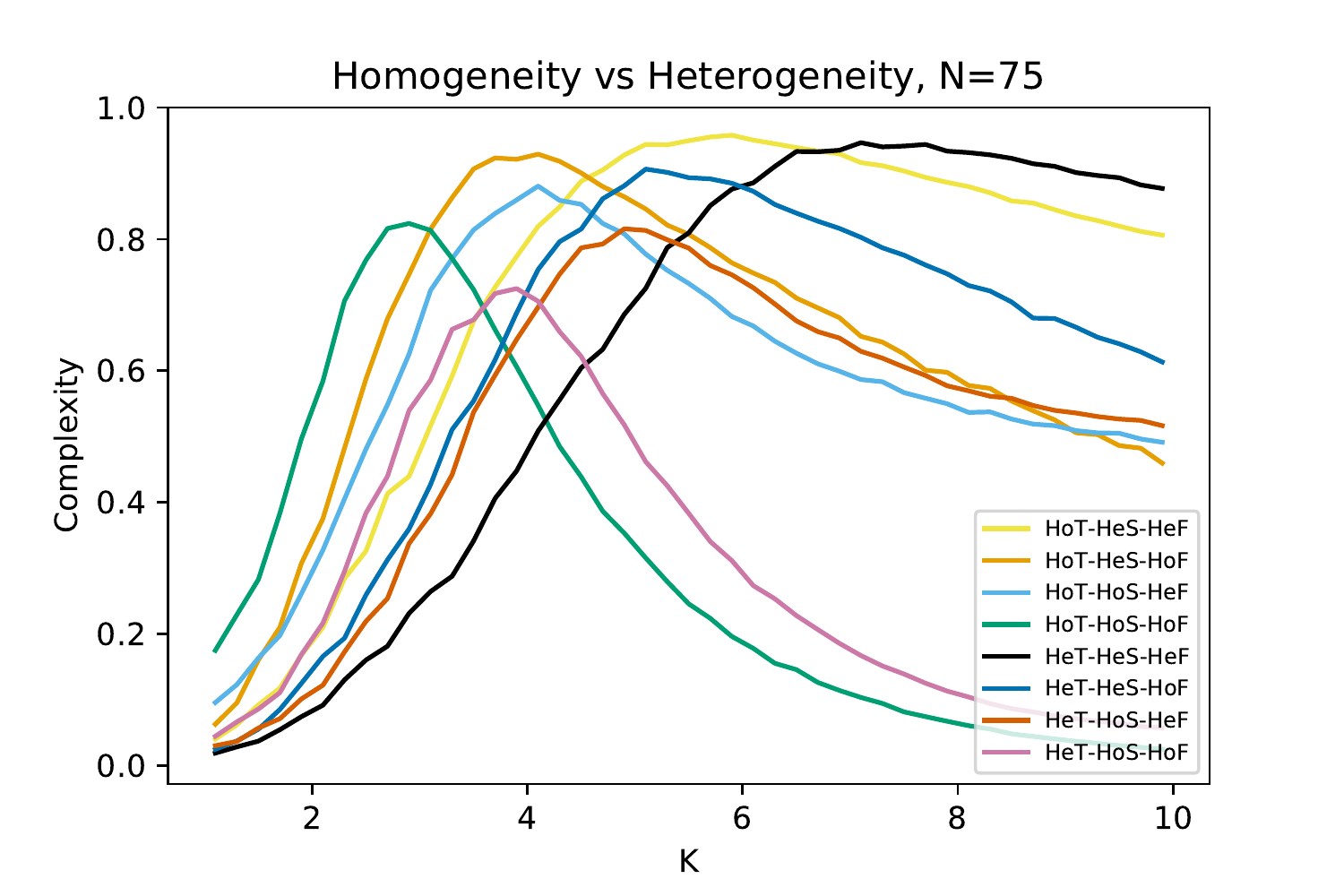} 
    \caption{} 
    \label{N75} 
    \vspace{4ex}
  \end{subfigure} 
  \begin{subfigure}[b]{0.5\linewidth}
    \centering
    \includegraphics[scale=0.5]{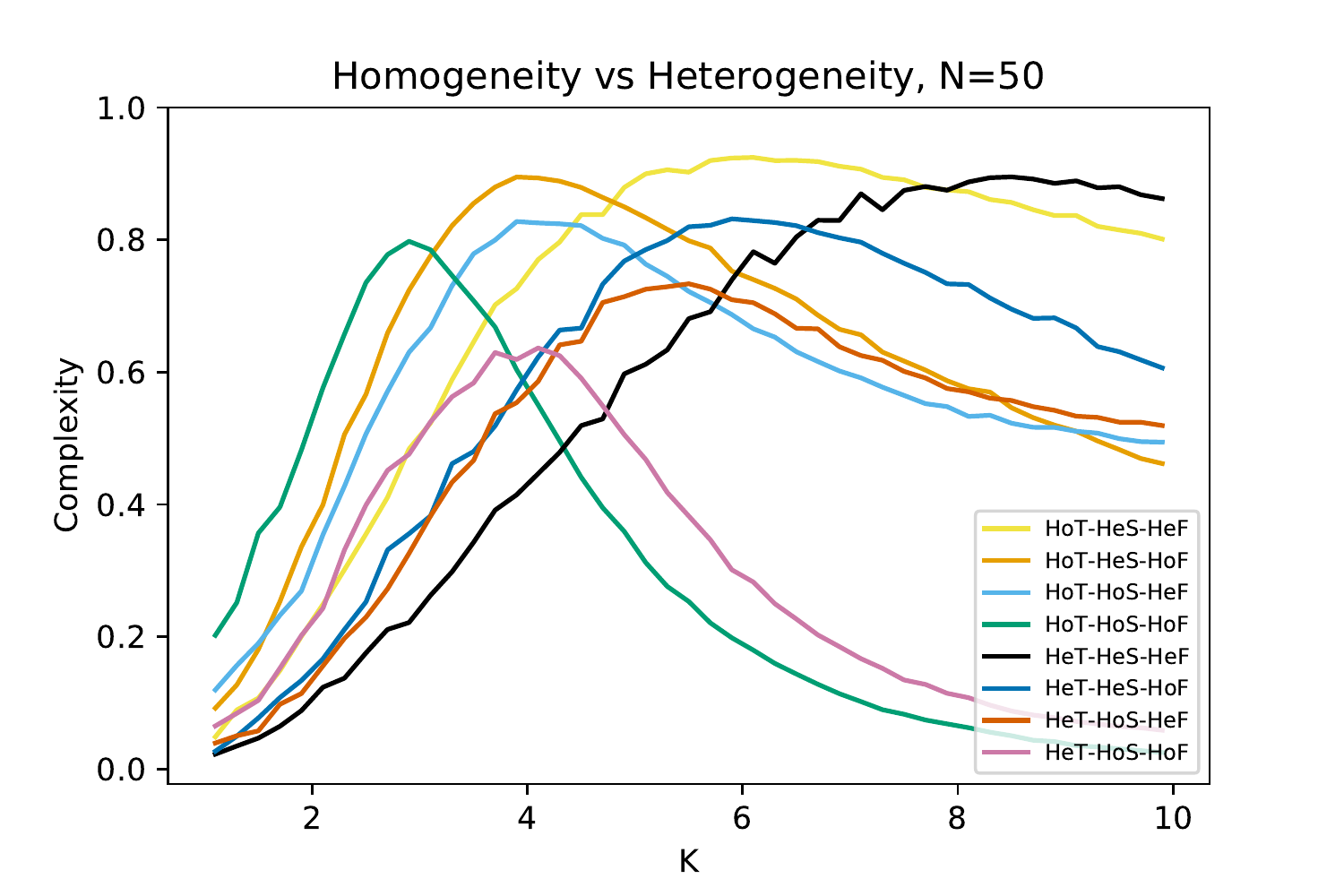} 
    \caption{} 
    \label{N50} 
  \end{subfigure}
   \begin{subfigure}[b]{0.5\linewidth}
    \centering
    \includegraphics[scale=0.5]{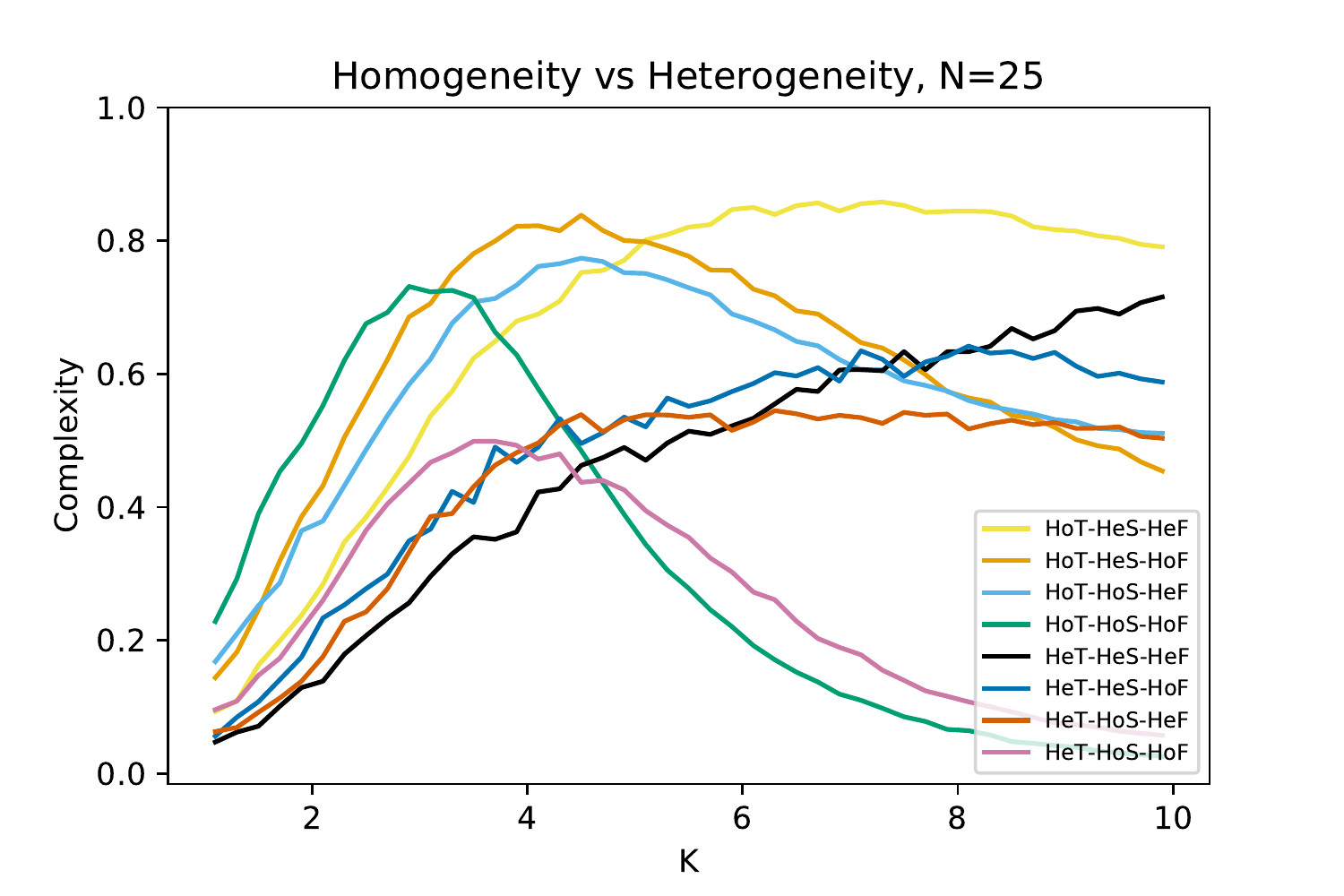} 
    \caption{} 
    \label{N25} 
  \end{subfigure}
  \caption{Average complexity $C$ of RBNs as the average connectivity $K$ is increased for the eight cases. The labels describe the combinations used; He stands for heterogeneity, Ho for homogeneity, T for temporal, S for structural and F for functional. The dissimilarities of these curves for different number of nodes in the network are shown. With $N=100$ nodes (\ref{N100}) it is observed that approximately from $K=6$ the triple heterogeneity (3He) dominates the other cases. For $N=75$ (\ref{N75}) the order of the curves is preserved but the value of $K$ needed to demarcate the dominance of 3He is higher, reaching a value close to $K=7$. With $N=50$ nodes (\ref{N50}) the curves look much flatter, despite the increase of noise in the curves the order of the curves is still preserved, but now an approximate value of $K=8$ is needed to visualize that 3He outperforms the other cases. At $N=25$ (\ref{N25}) the noise accumulates too much in the curves, however, by following the trend of the 3He curve it can be stated that the order will still be preserved, but now a much larger $K$ is required to display it. For all curves, a step of $\Delta K = 0.2$ was used, giving a total of $45$ $K$-connectivities. For each of these connectivity values, $1000$ networks were generated and their complexities were averaged.}
  \label{N}
\end{figure}
Figure~\ref{N} shows the complexity vs. connectivity curves for the eight cases which are the combinations among structural, temporal and functional homogeneity/heterogeneity. For the temporal heterogeneity, we are considering an out-degree updating. The distribution used for the functional heterogeneity is triangular with a mean of $0.5$ and its domain its the whole interval $[0,1]$. The curves that reach lower values of complexity represent the totally homogeneous case (HoSHoTHoF), while the curves that reach higher values of complexity are the threefold heterogeneity case (HeSHeTHeF). 

In CRBNs (3Ho), the size $N$ of the network does not seem to affect that much results, but this is not the case as heterogeneity is added.
The $K$ needed to ensure that the triple heterogeneity (3He) maximizes criticality depends on the number of nodes. For $N=100$ (\ref{N100}) the curves look smooth and the minimum $K$ at which 3He is greater than all cases is approximately $K=6$. In the case $N=75$ (\ref{N75}) the order of the curves is preserved with respect to the previous case, but the curves look flattened. Moreover, the $K$ required for 3He to be maximal is close to $K=7$. Although for $N=50$ (\ref{N50}) the order of the curves is also preserved, the $K$ required for 3He to be the upper curve is now a value close to $K=8$. Finally we observe that at $N=25$ (\ref{N25}) the curves are less smooth. Also, by the trend of the HeTHeSHeF (3He) curve, we can state that the hierarchy is still fulfilled but now a larger $K$ would be needed to better visualize it.

If to focus on a single type of heterogeneity, structural (HoTHeSHoF) leads to the highest increase of complexity $C$, followed by temporal (HeTHoSHoF), while functional (HoTHoSHeF) has lower maximum $C$ than 3Ho, but tends to have a higher $C$ for larger $K$.

\begin{figure}[htbp] 
  \begin{subfigure}[b]{0.5\linewidth}
    \centering
    \includegraphics[scale=0.4]{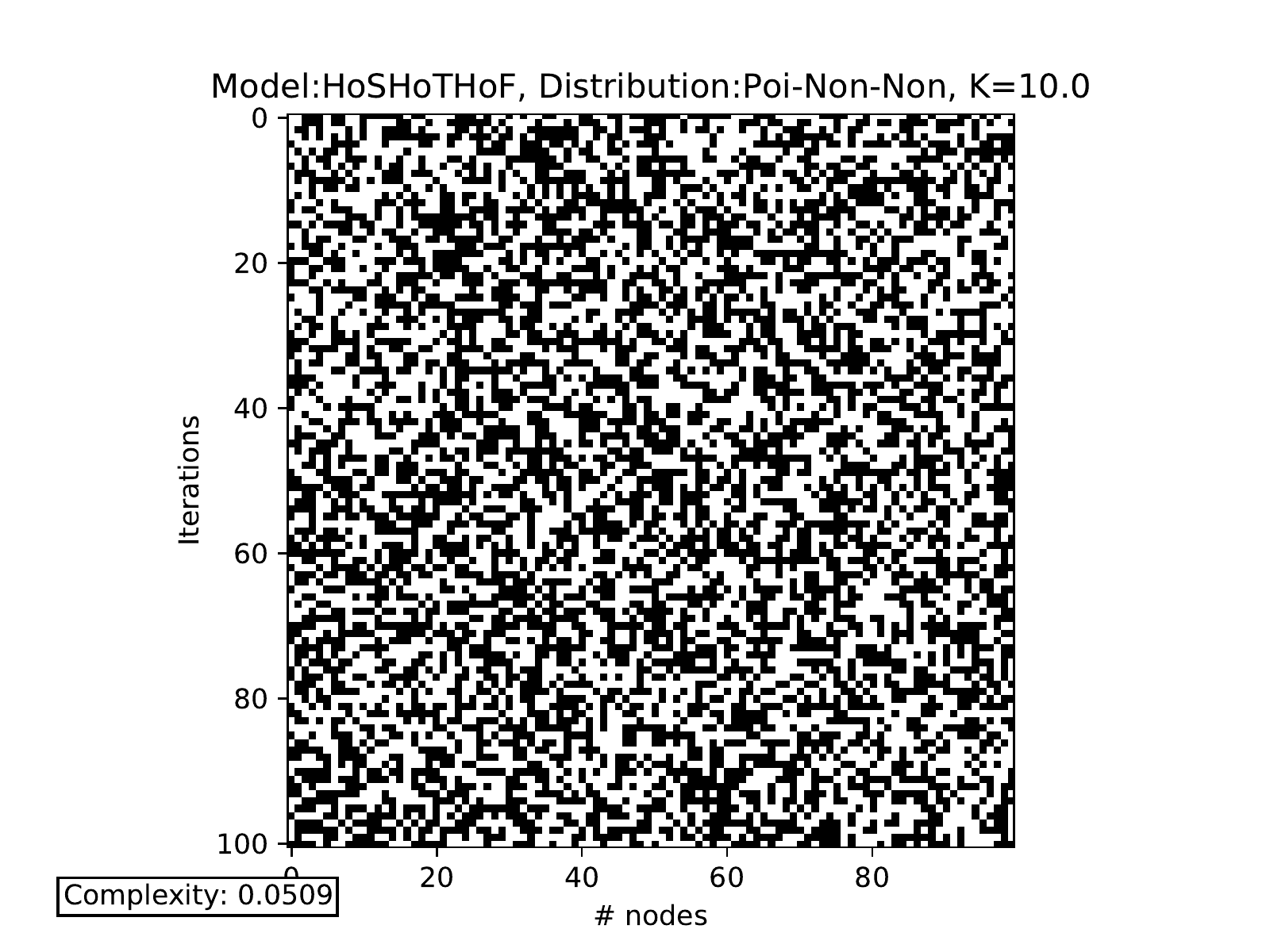} 
    \caption{} 
    \label{HoSHoTHoF} 
    \vspace{4ex}
  \end{subfigure}
  \begin{subfigure}[b]{0.5\linewidth}
    \centering
    \includegraphics[scale=0.4]{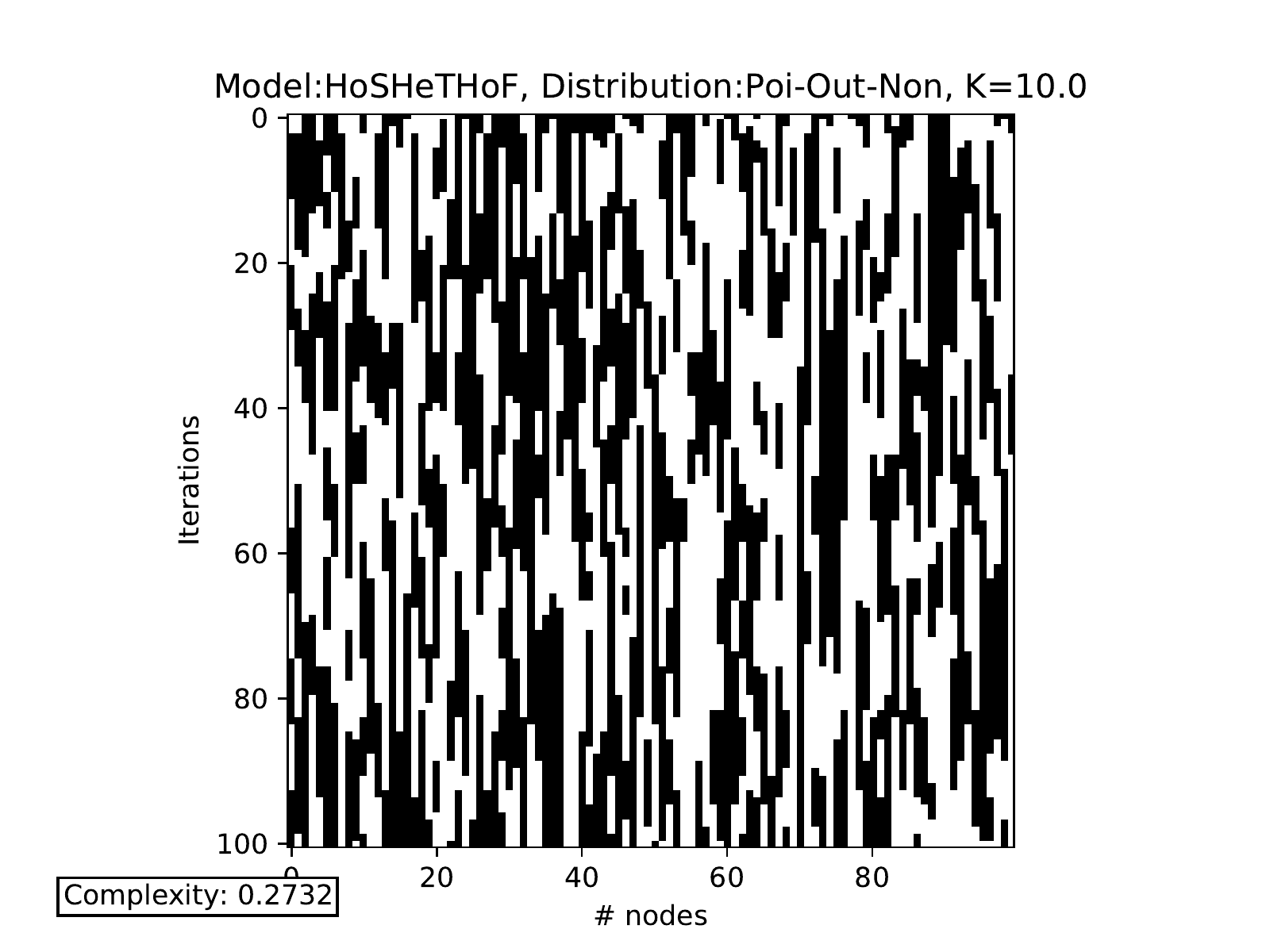} 
    \caption{} 
    \label{HoSHeTHoF} 
    \vspace{4ex}
  \end{subfigure} 
  \begin{subfigure}[b]{0.5\linewidth}
    \centering
    \includegraphics[scale=0.4]{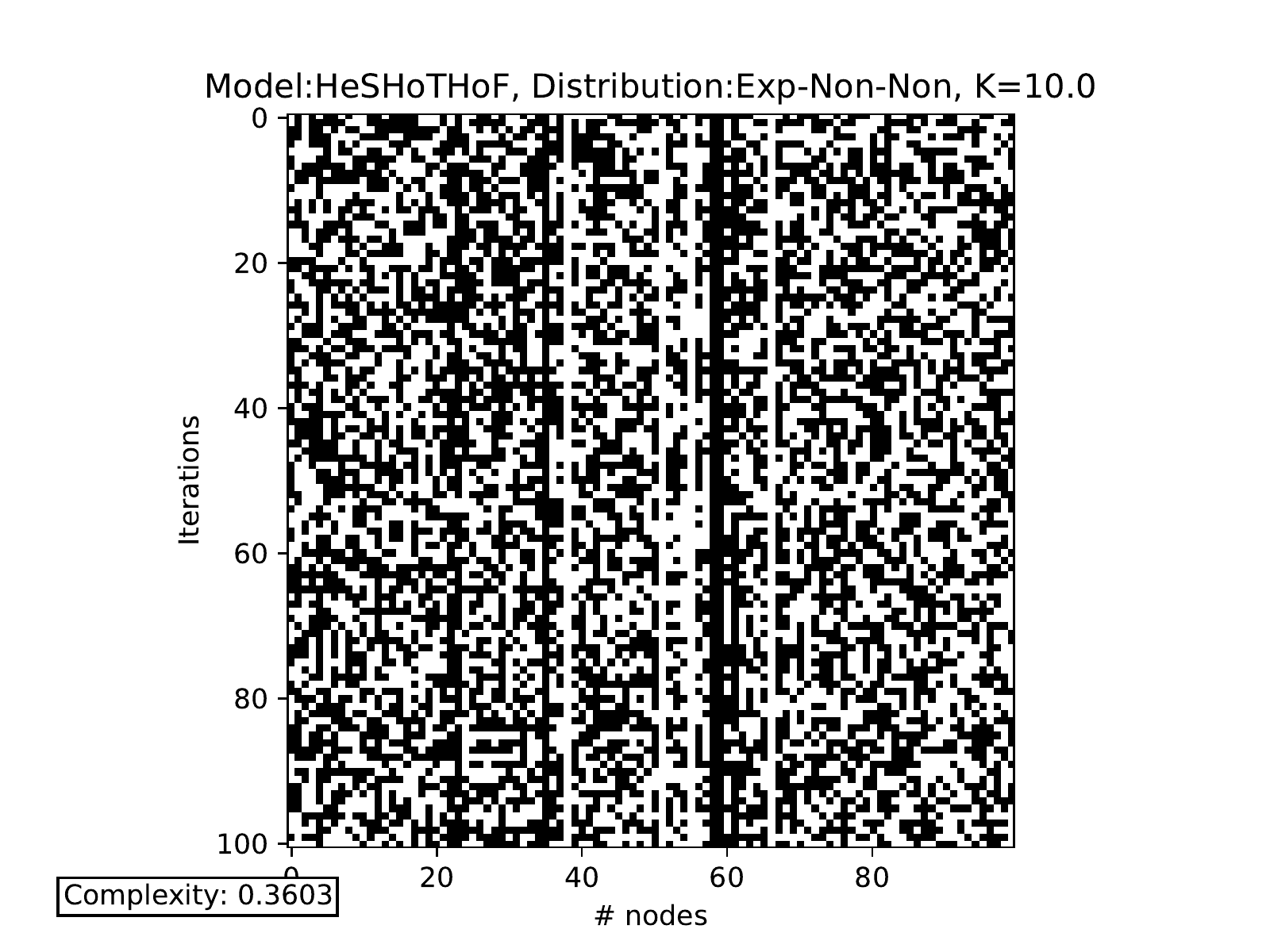} 
    \caption{} 
    \label{HeSHoTHoF} 
  \end{subfigure}
   \begin{subfigure}[b]{0.5\linewidth}
    \centering
    \includegraphics[scale=0.4]{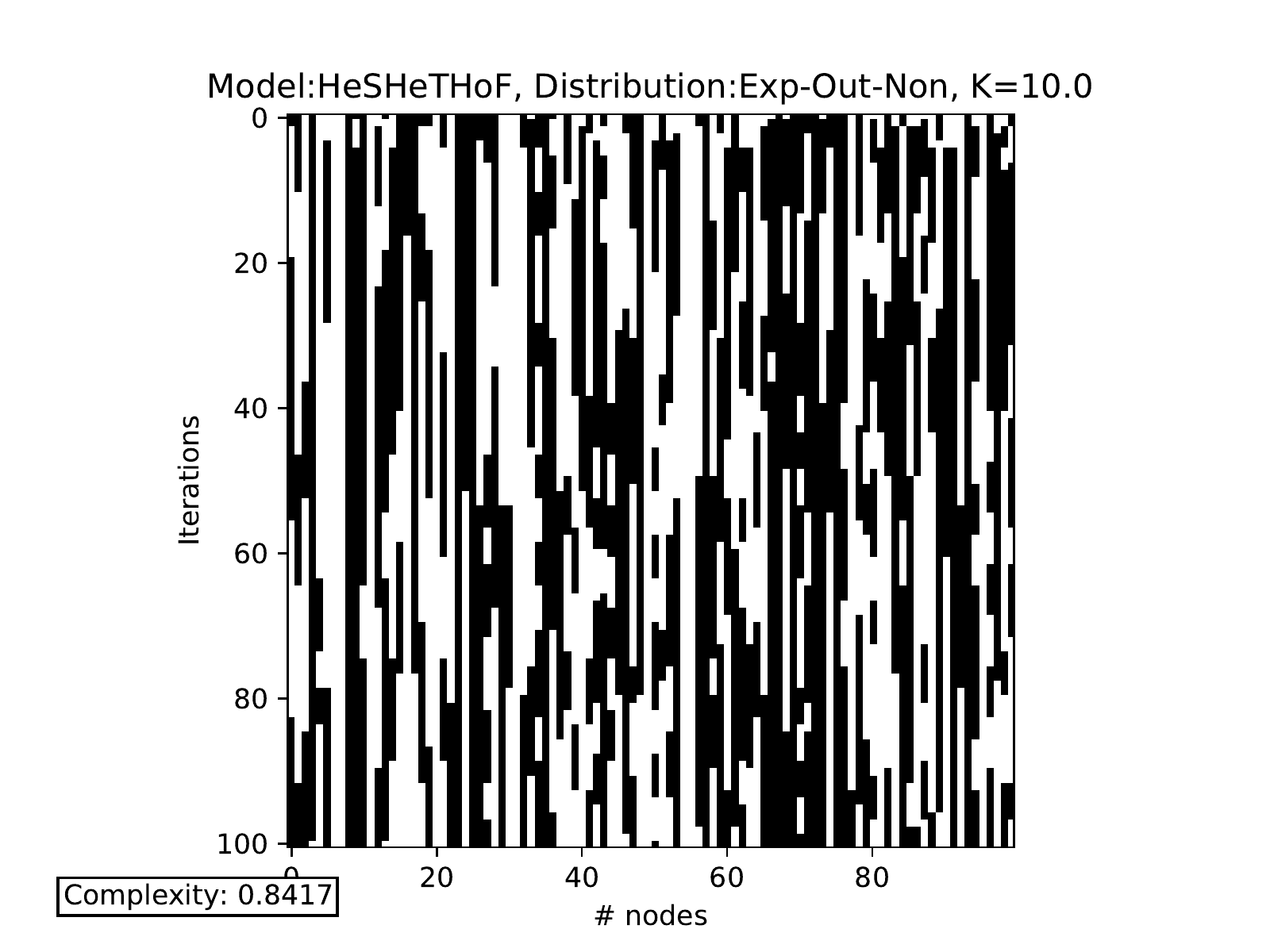} 
    \caption{} 
    \label{HeSHeTHoF} 
  \end{subfigure}
  \caption{State transitions by iteration for functional homogeneity. Time flows down (initial state is the top column). In the upper part of the subfigures we indicate the distributions used for each case. 'Poi' stands for Poisson, 'Exp' for Exponential, 'Out' for Out-degree and 'Non' for None. At the bottom of each subfigure the average complexity (per iteration) for each case is shown. The iterations for HoSHoTHoF (\ref{HoSHoTHoF}) resemble white noise, this coincides with the fact that the complexity obtained is very low. Adding only temporal heterogeneity (\ref{HoSHeTHoF}) gives a little more structure but the complexity is still very low. Adding only structural heterogeneity (\ref{HeSHoTHoF}) does not seem to gain as much structure in the iterations but still the complexity is higher than in HoSHeTHoF. Adding both temporal and structural heterogeneity (\ref{HeSHeTHoF}) gives more structure to the dynamics and the complexity increases more. To generate these images we consider a network with $100$ nodes, $100$ steps each (time flows downwards) and a connectivity of $K=10$.}
  \label{V1}
\end{figure}

\begin{figure}[htbp] 
  \begin{subfigure}[b]{0.5\linewidth}
    \centering
    \includegraphics[scale=0.4]{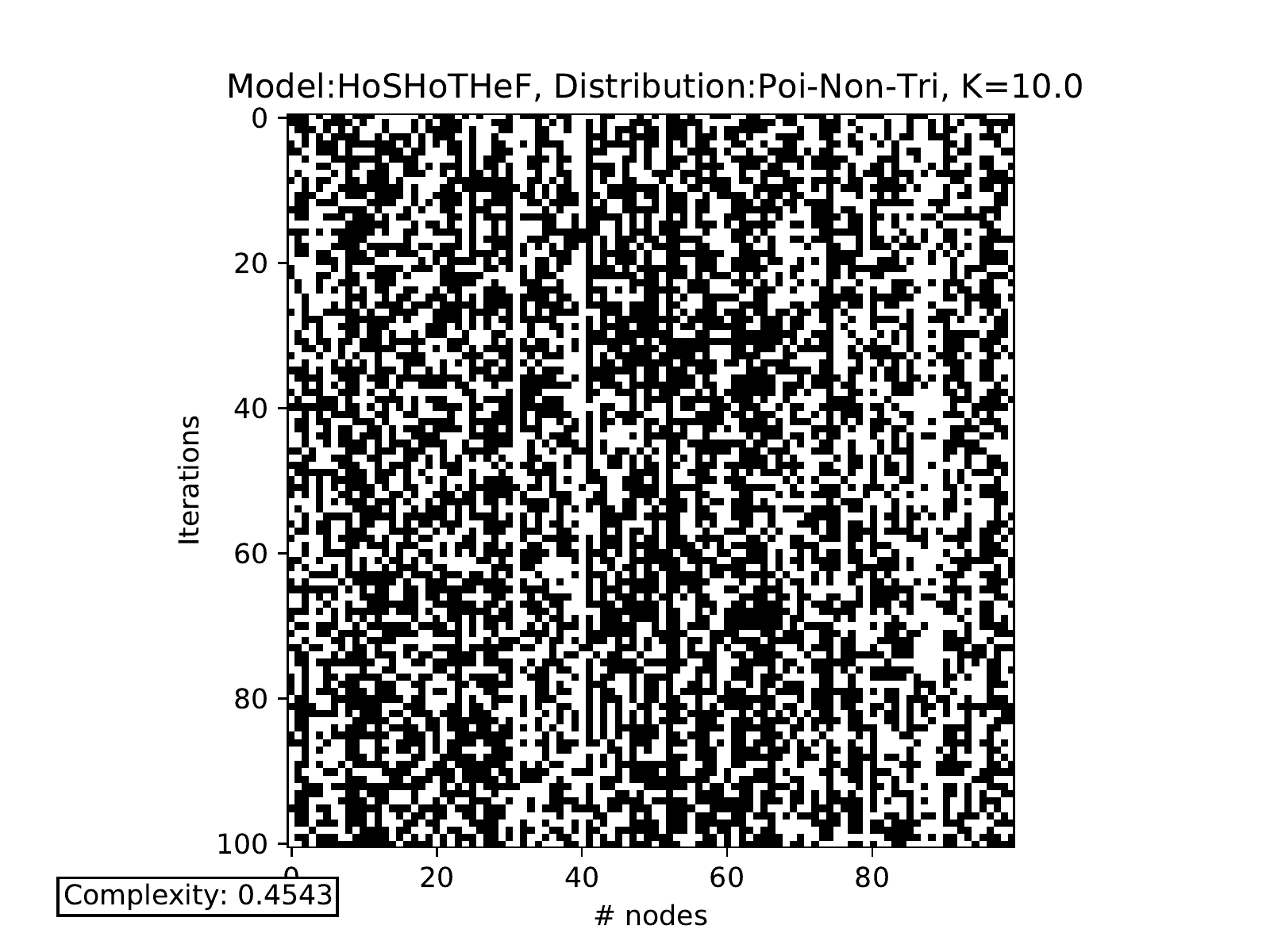} 
    \caption{} 
    \label{HoSHoTHeF} 
    \vspace{4ex}
  \end{subfigure}
  \begin{subfigure}[b]{0.5\linewidth}
    \centering
    \includegraphics[scale=0.4]{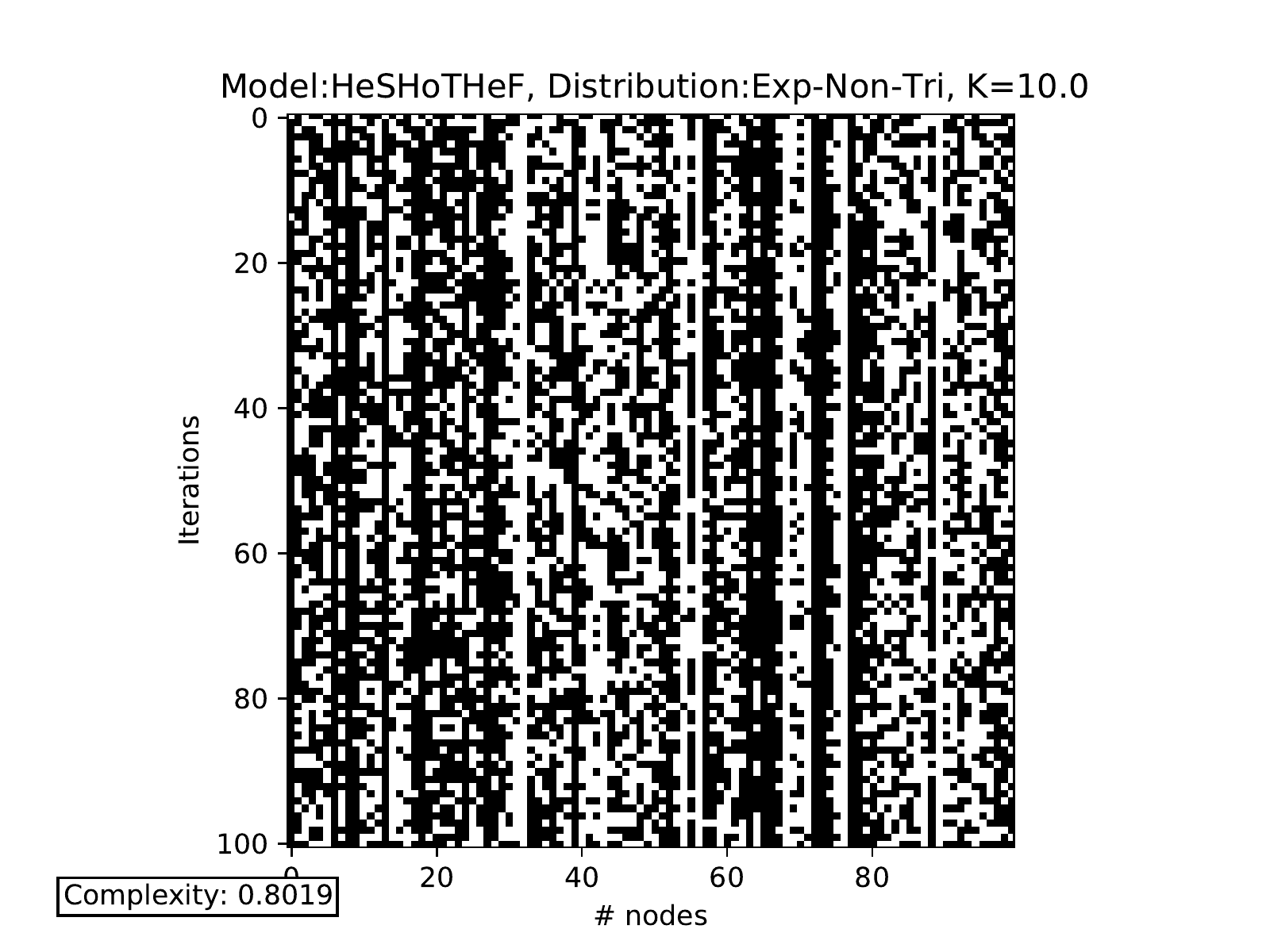} 
    \caption{} 
    \label{HeSHoTHeF} 
    \vspace{4ex}
  \end{subfigure} 
  \begin{subfigure}[b]{0.5\linewidth}
    \centering
    \includegraphics[scale=0.4]{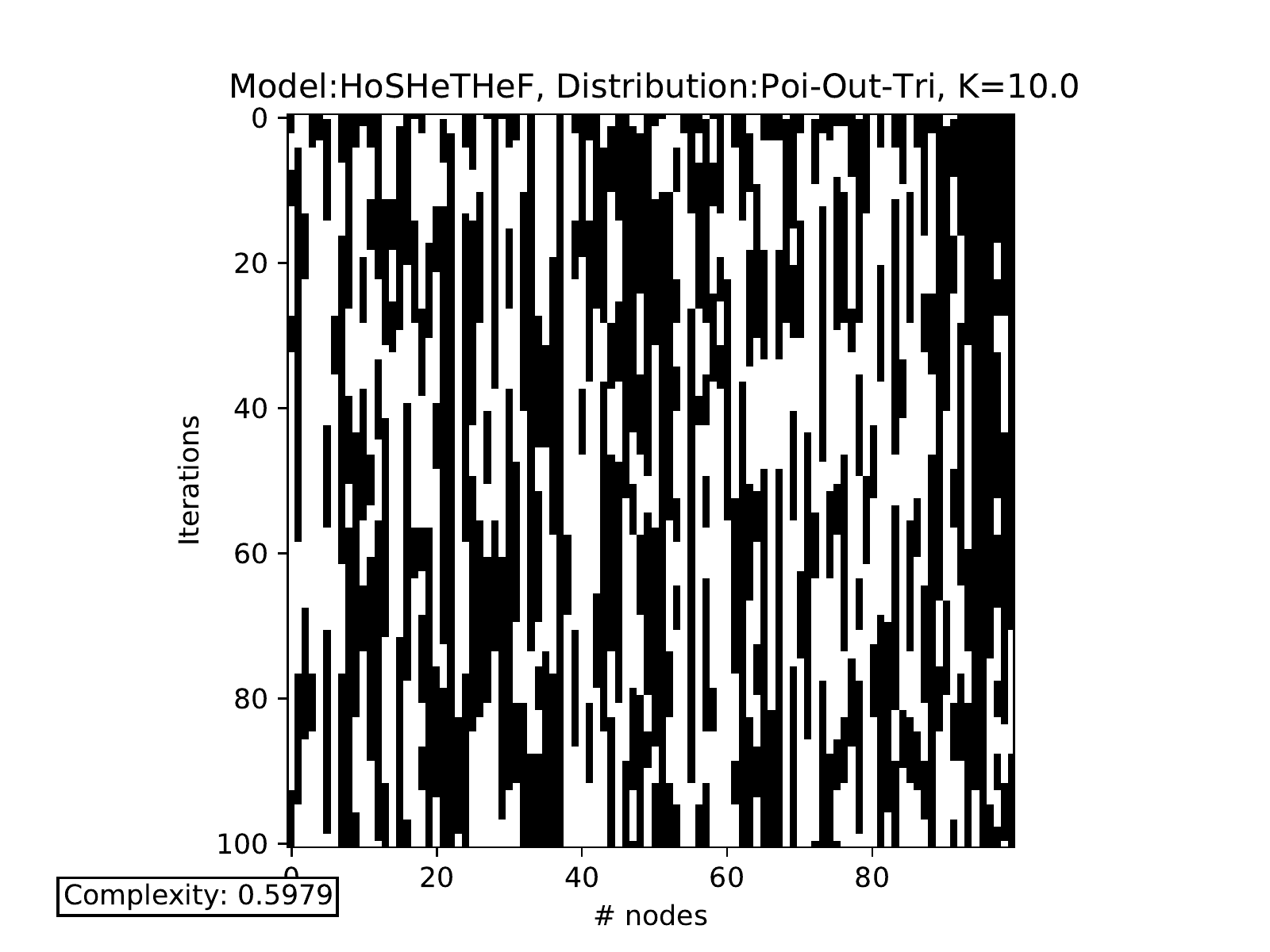} 
    \caption{} 
    \label{HoSHeTHeF} 
  \end{subfigure}
   \begin{subfigure}[b]{0.5\linewidth}
    \centering
    \includegraphics[scale=0.4]{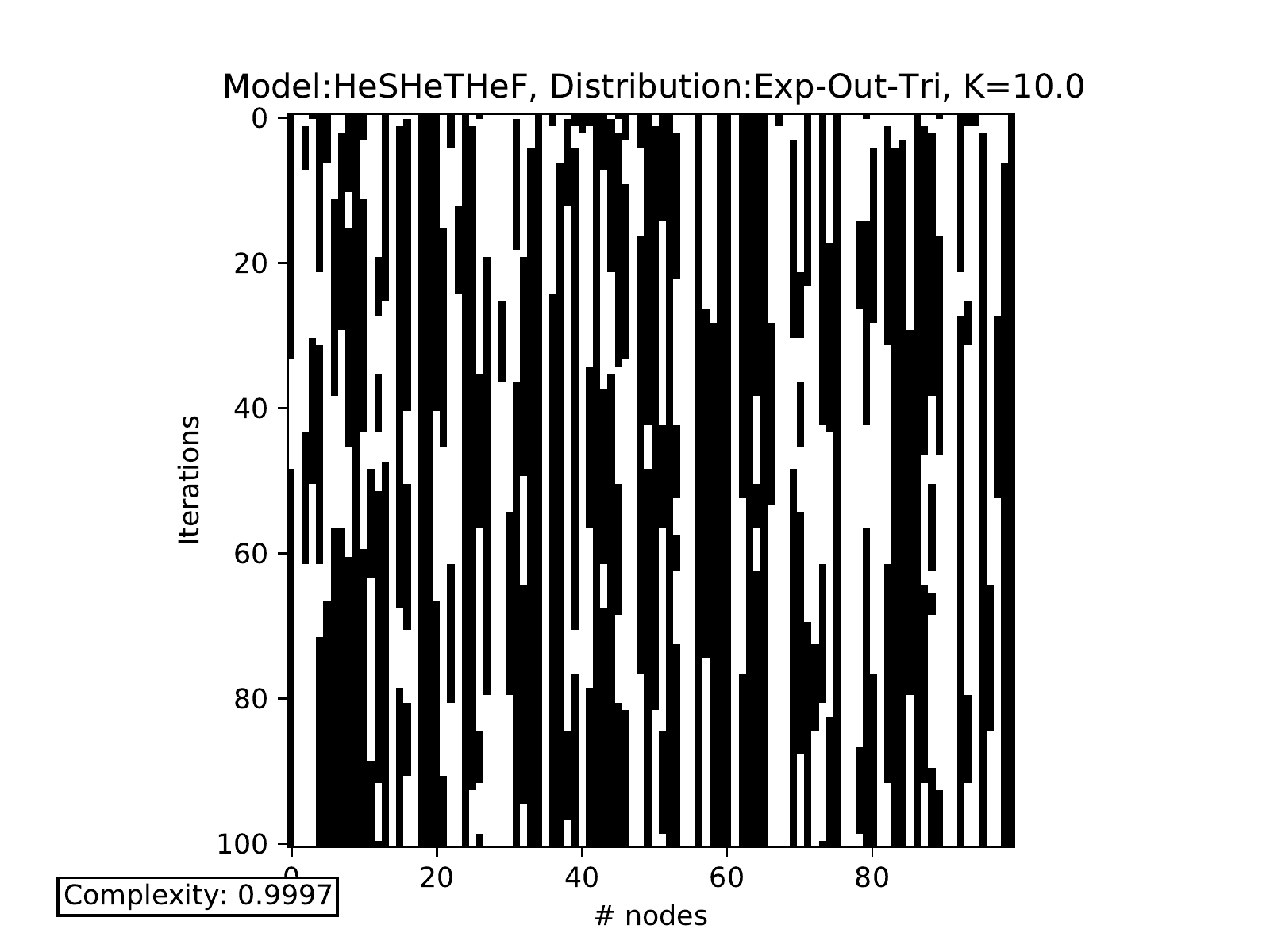} 
    \caption{} 
    \label{HeSHeTHeF} 
  \end{subfigure}
  \caption{State transitions by iteration for functional heterogeneity. In the upper part of the subfigures we indicate the distributions used for each case. 'Poi' stands for Poisson, 'Exp' for Exponential, 'Out' for Out-degree, 'Tri' for Triangular and 'Non' for None. At the bottom of each subfigure the average complexity (per iteration) for each case is shown. When only functional heterogeneity is added (\ref{HoSHoTHeF}), greater complexity is observed with respect to HoSHeTHoF and HeSHoTHoF. Adding functional and structural heterogeneity (\ref{HeSHoTHeF}) results in similar complexity to the HeSHeTHoF case, but having functional and temporal heterogeneity (\ref{HoSHeTHeF}) keeps the complexity at an intermediate value. This is consistent with what we have seen previously, HeS increases the complexity more than HeT. Finally, HeSHeTHeF (\ref{HeSHeTHeF}) is the case that gives the major structure to the iterations. In fact, complexity is close to maximal for this case. To generate these images we consider a network with $100$ nodes, $100$ steps each (time flows downwards) and a connectivity of $K=10$.}
  \label{V2}
\end{figure}

As the number of nodes ($N$) increases, the minimum connectivity ($K$) for the triple heterogeneity to be beneficial decreases. Thus, there is a relationship between the number of nodes in the network and the critical connectivity $K_c$ for heterogeneous RBNs. 
Furthermore, heterogeneity seems to be additive (the more types of heterogeneity, the more criticality). Although for very large $K$ we could claim that the 3He maximally extends criticality, we run into different limitations. The first is algorithmic, since obtaining the complexity vs connectivity curves for larger $K$ is computationally expensive. This implies we are limited to lower ranges of $K$. Since the 3He curve rises for higher $K$, then the area under the $C$ curve (a measure of criticality that could be used) for other cases is larger than for 3He. Even if we try to truncate these complexity curves, it may be 3He is not the highest $C$ for all values of $N$. Therefore, it is necessary to find another way to prove that the triple heterogeneity extends criticality the most. Moreover, the fact that we need a very large $K$ to obtain maximum criticality given a specific configuration implies that if a complex system is under certain conditions (e.g., suppose resources are scarce or limited) then a certain amount of homogeneity will be better in order to achieve higher criticality. In the view of the mentioned above and to give a qualitative proof with the advantages of the triple heterogeneity case, for each combination we present in Figures~\ref{V1} and~\ref{V2} the transitions of the states of the nodes in the network for a fixed $K$.

The transitions for the functionally homogeneous cases ($N=100, K=10$) are shown in Figure~\ref{V1}. In HoSHoTHoF (\ref{HoSHoTHoF}) we observe a clear similarity with white noise. In fact, the complexity for this case is lower than for the other cases. From HoSHeTHoF (\ref{HoSHeTHoF}) and HeSHoTHoF (\ref{HeSHoTHoF}) we can state that in this scenario structural heterogeneity (HeS) contributes a greater extent of criticality than temporal heterogeneity (HeT). This is despite the fact that HeT seems to give much more structure to the HoSHoTHoF white noise. Finally, by combining the effect of HeS and HeT we obtain HeSHeTHoF (\ref{HeSHeTHoF}) and this results in transitions with more structure. Note how the complexity of this last case ($C=0.8417$) is larger than the sum of the two previous cases ($C = 0.2732 + 0.3603 = 0.6335$). The above suggests that heterogeneity is only qualitatively (not numerically) additive; the greater the heterogeneity, the greater the complexity (and hence, the more likely a higher criticality threshold). 

The dynamic transitions for the functionally heterogeneous cases are shown in Figure~\ref{V2} (also $N=100, K=10$). The notation is essentially the same as in the previous case, with the difference that we now add 'Tri' to refer to the use of the triangular distribution in the functional case. A higher complexity value is observed in HoSHoTHeF (\ref{HoSHoTHeF}) than in HoSHeTHoF (\ref{HoSHeTHoF}) and HeSHoTHoF (\ref{HeSHoTHoF}), this implies that functional heterogeneity (HeF) outperforms HeS and HeT in extending criticality separately (remember that we are focussing only on $K=10$). However, HeF alone does not outperform the sum of HeS and HeT. On the other hand, both HeSHoTHeF (\ref{HeSHoTHeF}) and HoSHeTHeF (\ref{HoSHeTHeF}) are consistent with previous observations, in which HeS provides a greater extension of criticality with respect to HeT; to verify the above it suffices to compare the complexities of these two cases. Finally, we observe that 3He (\ref{HeSHeTHeF}) has higher complexity with respect to all the previous cases, confirming once again that the more heterogeneities, the higher the complexity. It is important to mention that these calculations are for a relatively large $K$. However, for smaller $K$ (and different values of $N$) the above is not always true. In addition we should also note that it is not true that the sum of the complexities for HeSHoTHoF (\ref{HeSHoTHoF}), HoSHeTHoF (\ref{HoSHeTHoF}) and HoSHoTHeF (\ref{HoSHoTHeF}) is equal to the complexity produced by HeSHeTHeF (\ref{HeSHeTHeF}). Again, heterogeneity is not numerically additive. 

\subsection{Varying functional and temporal parameters}
Now that we know the effects of heterogeneous functionality in extending criticality, we can explore the parameter space associated with HeSHeTHeF and study which variations further extend criticality.

\begin{figure}[htbp] 
  \begin{subfigure}[b]{0.5\linewidth}
    \centering
    \includegraphics[scale=0.5]{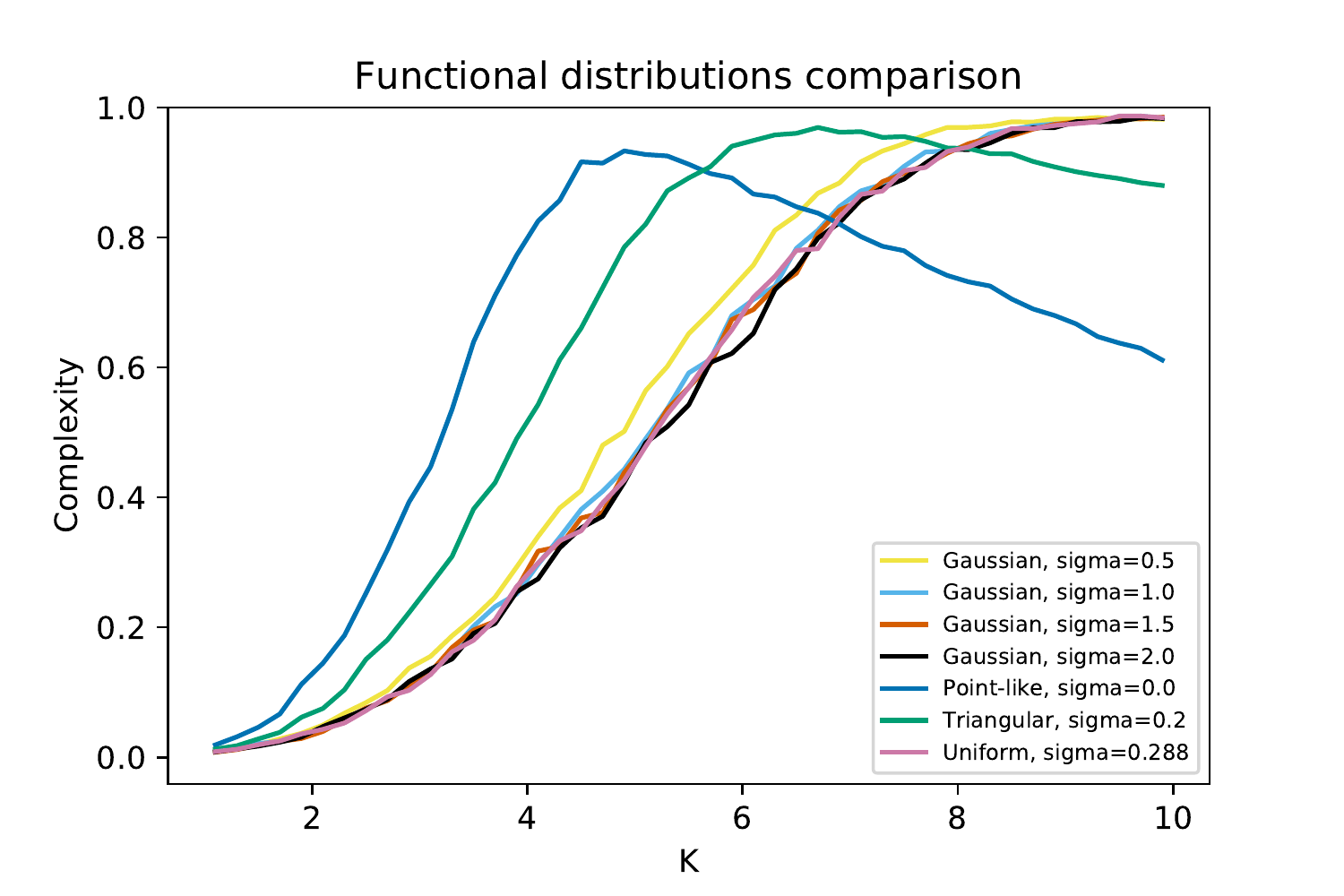} 
    \caption{ } 
    \label{FD} 
    \vspace{4ex}
  \end{subfigure}
  \begin{subfigure}[b]{0.5\linewidth}
    \centering
    \includegraphics[scale=0.5]{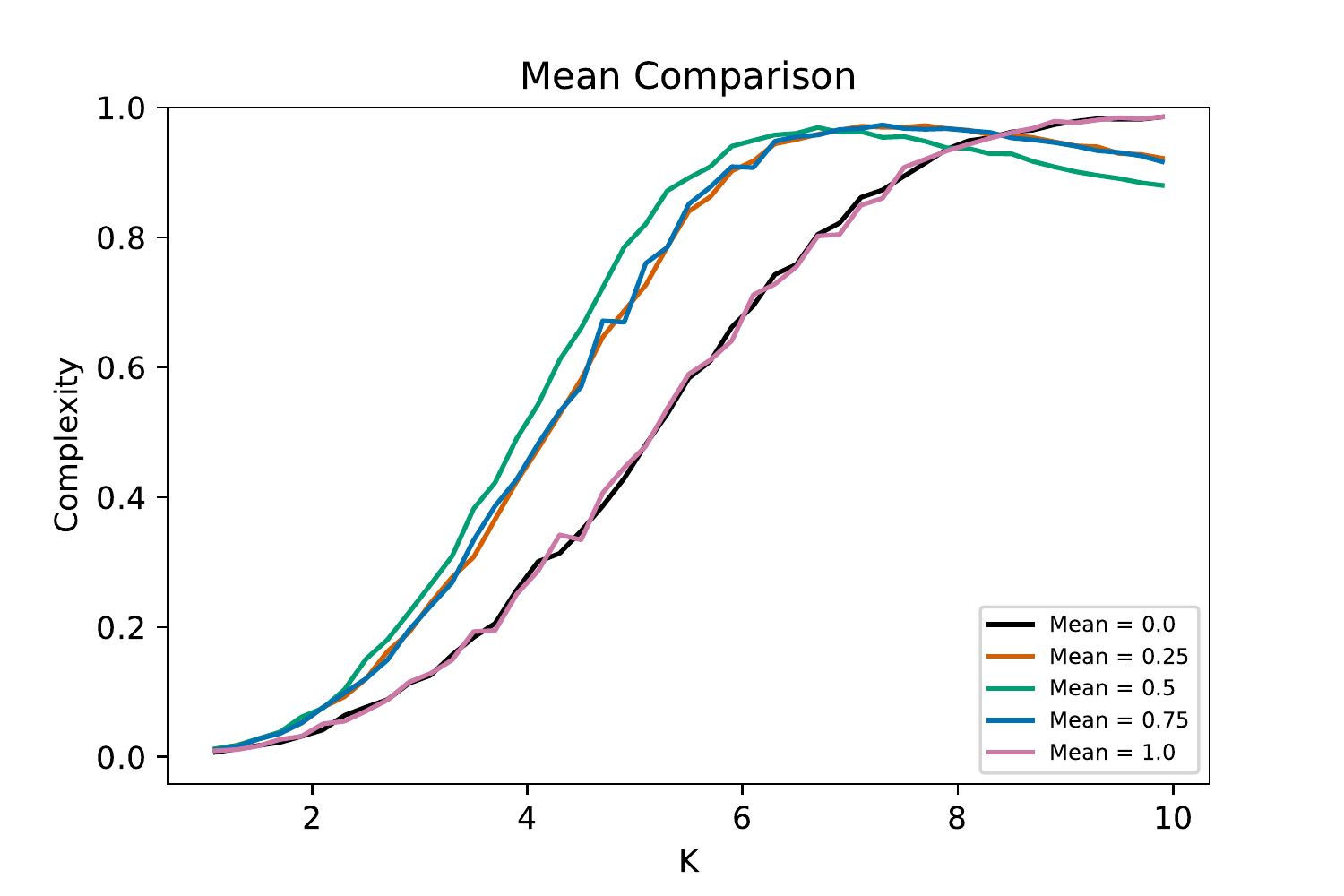} 
    \caption{ } 
    \label{Mean} 
    \vspace{4ex}
  \end{subfigure} 
  \begin{subfigure}[b]{0.5\linewidth}
    \centering
    \includegraphics[scale=0.5]{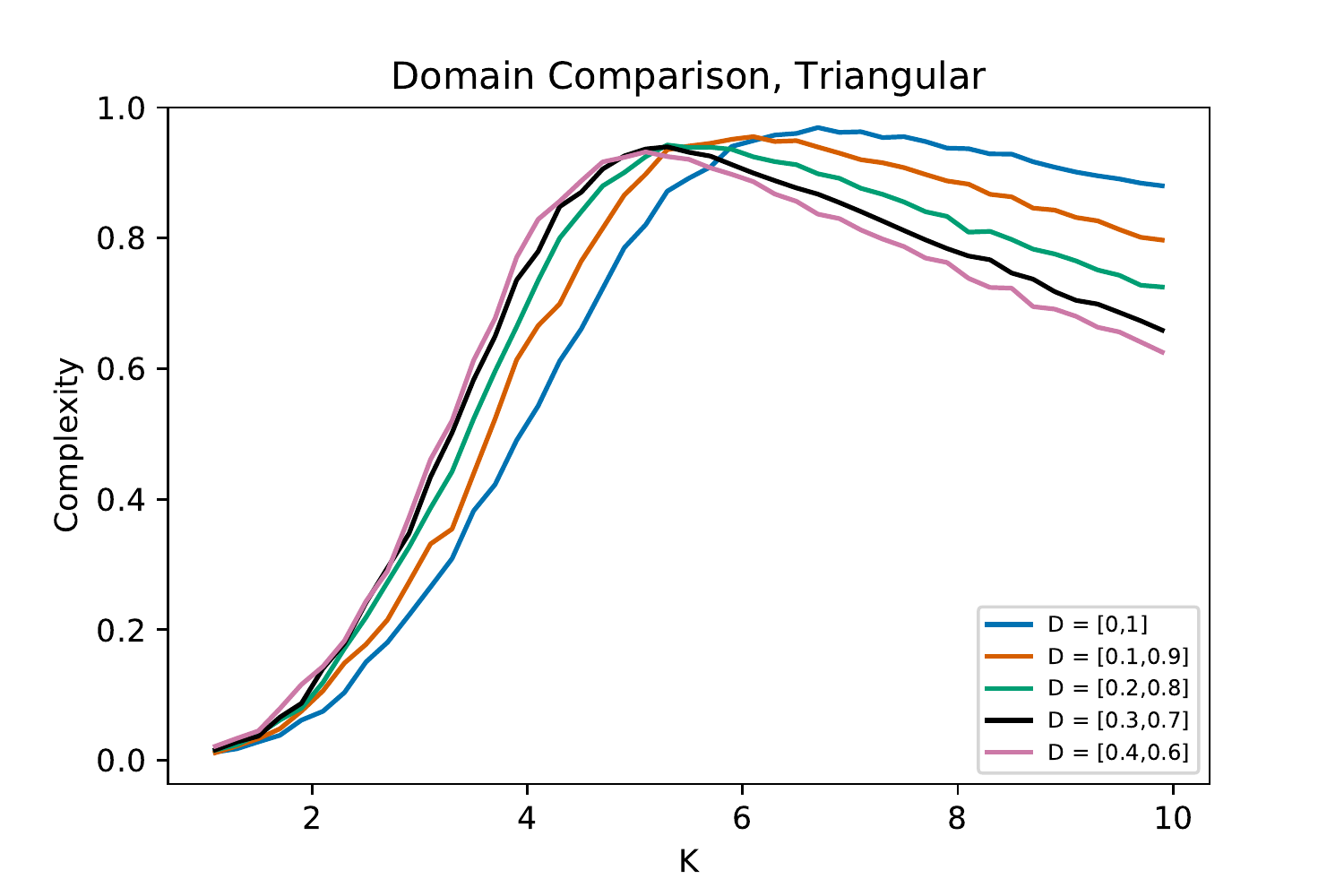} 
    \caption{ } 
    \label{TriD} 
  \end{subfigure}
   \begin{subfigure}[b]{0.5\linewidth}
    \centering
    \includegraphics[scale=0.5]{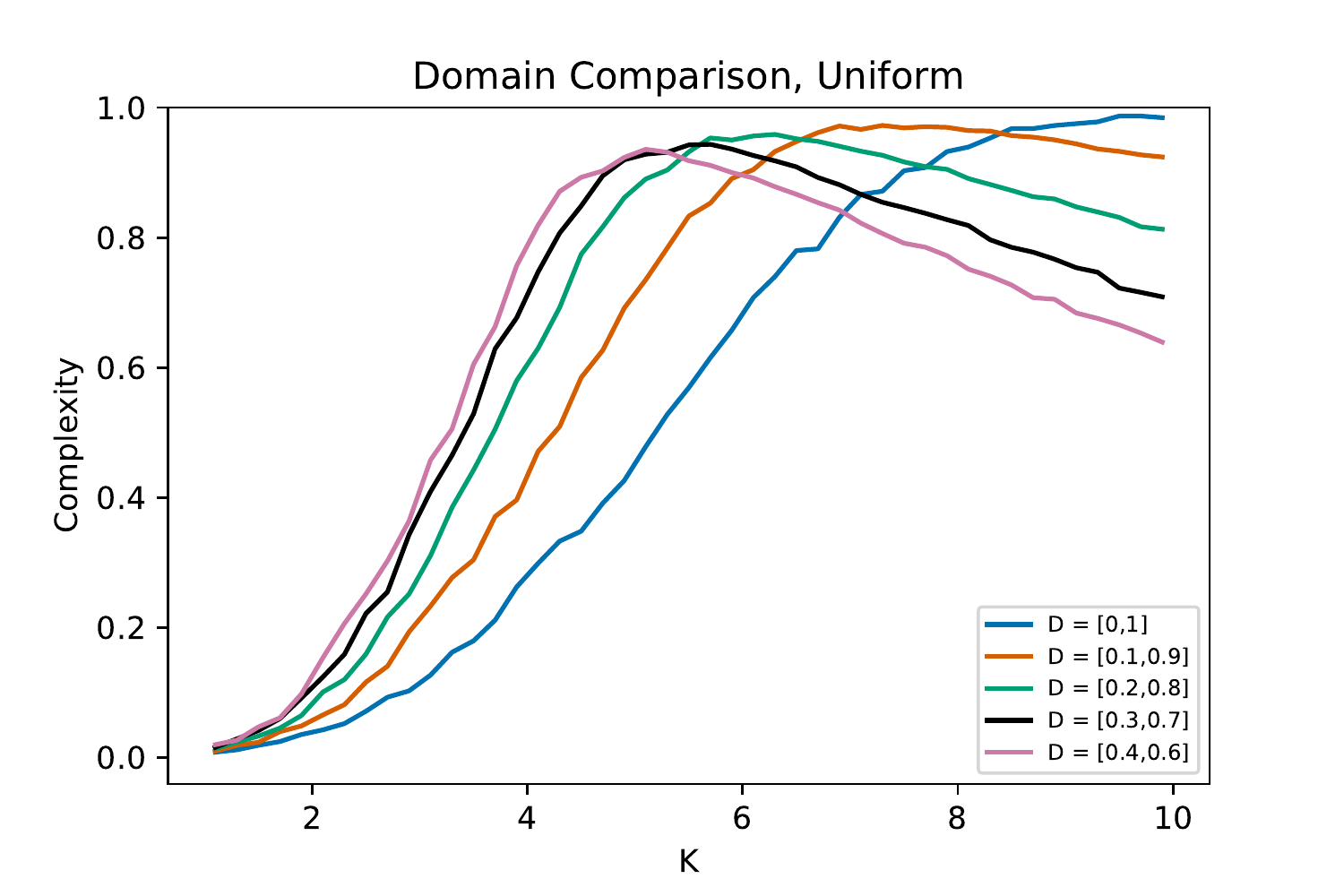} 
    \caption{ } 
    \label{UniD} 
  \end{subfigure}
  \caption{Average complexity of RBNs as the average connectivity $K$ is increased for variations in the functional parameters of the triple heterogeneity (3He) case. The subfigure~\ref{FD} shows the changes obtained by testing seven different functional distributions. Point-like indicates the extreme case when the standard deviation is zero, this is equivalent to the homogeneous functional case. When the distribution is Gaussian, convergence is observed after a certain value for the standard deviation. This behavior is not generalizable, because with the uniform distribution something similar is obtained and its standard deviation is smaller. By varying only the mean of the functional distribution (\ref{Mean}) we observe symmetric behavior, so that varying the mean only matters how far we are from $\mu=0.5$.~\ref{TriD} and~\ref{UniD} show what happens when the domain of the functional distribution is modified. In both cases, a similar order in the curves is observed and also as we make the interval shorter the curves stick together. For all curves, a step of $\Delta K = 0.2$ was used and in every step a total of 1000 iterations were averaged.}
  \label{Functional}
\end{figure}
Figure~\ref{FD} shows the complexity curves for seven different functional distributions. Four of them are Gaussian but with different standard deviations. The extreme case corresponds to PL (abbreviation of Point-like distribution) which is equivalent to the homogeneous functional case and has zero standard deviation. Approximately for values greater than $K=6.5$ the Gaussian and uniform distributions overcome the complexity of the extreme case. However, this is not the case for the triangular distribution, since for it a much higher $K$ is needed to extend the criticality. Since working with larger $K$ values is computationally expensive, it is not possible at this time to know whether the sigmoid-like curves will continue to grow or at some point begin to decrease. It is clear that for Gaussian distributions there is a value of $\sigma$ (close to $0.5$) such that after this number the result no longer depends on the chosen standard deviation. This property of the standard deviations cannot be generalized, for example, the uniform distribution has $\sigma = 0.288$ which is much smaller than the values of $\sigma$ for the Gaussian distributions and still has a behavior very similar to them. Finally we can state that qualitatively both the triangular and PL cases have a larger area under the curve than the other cases (which seem to extend criticality better), this is another indication that total heterogeneity is not always better.

Figure~\ref{Mean} shows the complexity curves for different means in the triangular distribution. The extreme values in this case are $\mu = 0$ and $\mu = 1$ and both again produce sigmoidal-like behavior. It is interesting that both cases produce such similar curves and these are no exception, for $\mu=0.25$ and $\mu=0.75$ the curves also overlap, so we can say that there is a symmetry with respect to $\mu=0.5$. This is because of the complementarity of 0 and 1 given by the logical negation. Thus, choosing a different value for $\mu$ only matters how far away we are from $\mu=0.5$. Again, as we cannot reach higher values of $K$, we can only state that for this interval the smallest areas under the curve correspond to the extreme cases. This is due to the fact that they take longer to rise with respect to the other cases. Figure~\ref{TriD} shows the complexity curves taking different intervals as the domain for the triangular distribution. All intervals are symmetrical with respect to $0.5$ and as the interval length is reduced, the associated standard deviation also decreases. Before $K \approx 6$ all other curves exceed the one associated with the domain $D=[0,1]$. However, after $K \approx 6$ this curve is the one that surpasses the others. Because of this, it is difficult to know qualitatively which extends criticality more. However, it is clear that as we make the interval smaller, the curves stick together. This effect is even more visible in figure~\ref{UniD}, which is the same experiment but using a uniform distribution. To make the analysis a little more quantitative, a simple integration algorithm was implemented to calculate the area under the curves. By doing the above, it was obtained that for the triangular distribution the interval that extends better the criticality is $D=[0.1,0.9]$, while for the uniform distribution the interval that extends better the criticality is $D=[0.2,0.8]$. This implies that criticality depends not only on the length of the interval but also on the distribution used.

\begin{figure}[ht] 
        \centering \includegraphics[scale=0.7]{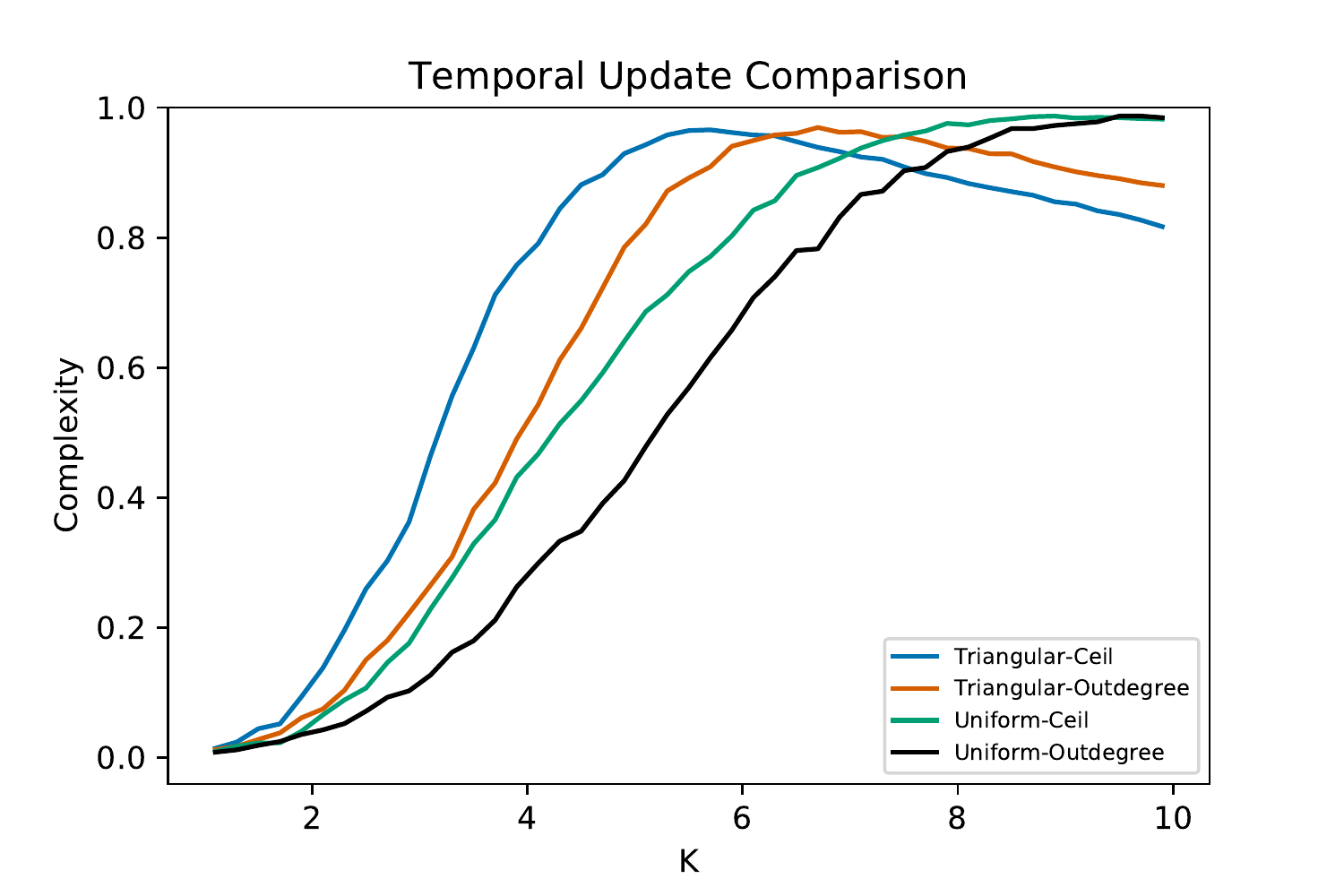}
        \caption{Average complexity of RBNs as the average connectivity $K$ is increased for two temporal strategies using two functional distributions. For the triangular distribution, Out-degree achieves a higher complexity for relatively large $K$. However, it is clear that for most values of $K$ the Ceil strategy is dominant. For the uniform distribution, Ceil is always superior to Out-degree. For very large values of $K$ a tie is observed but Out-degree never seems to completely outperform Ceil. For all curves, a step of $\Delta K = 0.2$ was used and in every step a total of 1000 iterations were averaged.}
        \label{TUp}
\end{figure}

So far, we have modified only the functional parameters. But to further complement this work, we will perform an experiment using two different temporal strategies. The first strategy is the Out-degree that we have been using along the paper and consist in “the more connections a node has, the slower it is updated”. The second strategy is called 'Ceil' and this assigns to all the nodes the same maximum activation period. In order to have a more exhaustive trial we decided to explore two types of functional distribution: the triangular and the uniform, both with a mean of $0.5$ and domain $[0,1]$. Figure~\ref{TUp} shows the results of the experiment described above. For both functional distributions, the Ceil strategy extends the criticality more compared to the Out-degree strategy we have been using. Furthermore, it appears that only for very large $K$ we could achieve an advantage by using Out-degree updating. However, as previously discussed, this is not always possible. In general terms, we could argue that Out-degree is a more heterogeneous temporal strategy compared to Ceil. However, it is the latter that gives us more criticality under a resource constrained scenario, which again shows that maximum heterogeneity should not always be preferred. 

\subsection{Antifragility}

\begin{figure}[ht] 
  \begin{subfigure}[b]{0.5\linewidth}
    \centering
    \includegraphics[scale=0.5]{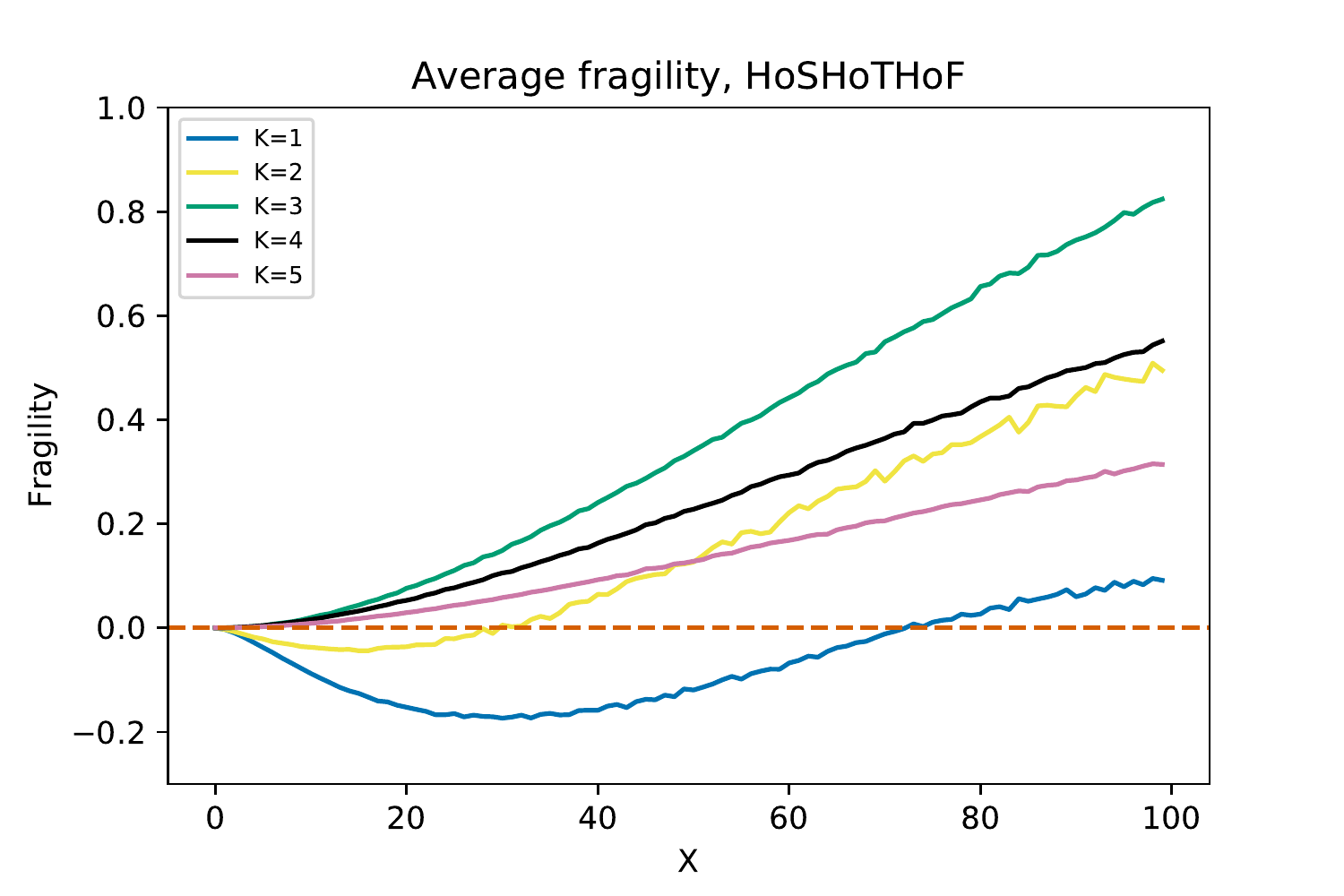} 
    \caption{ } 
    \label{AvsX_Ho} 
    \vspace{4ex}
  \end{subfigure}
  \begin{subfigure}[b]{0.5\linewidth}
    \centering
    \includegraphics[scale=0.5]{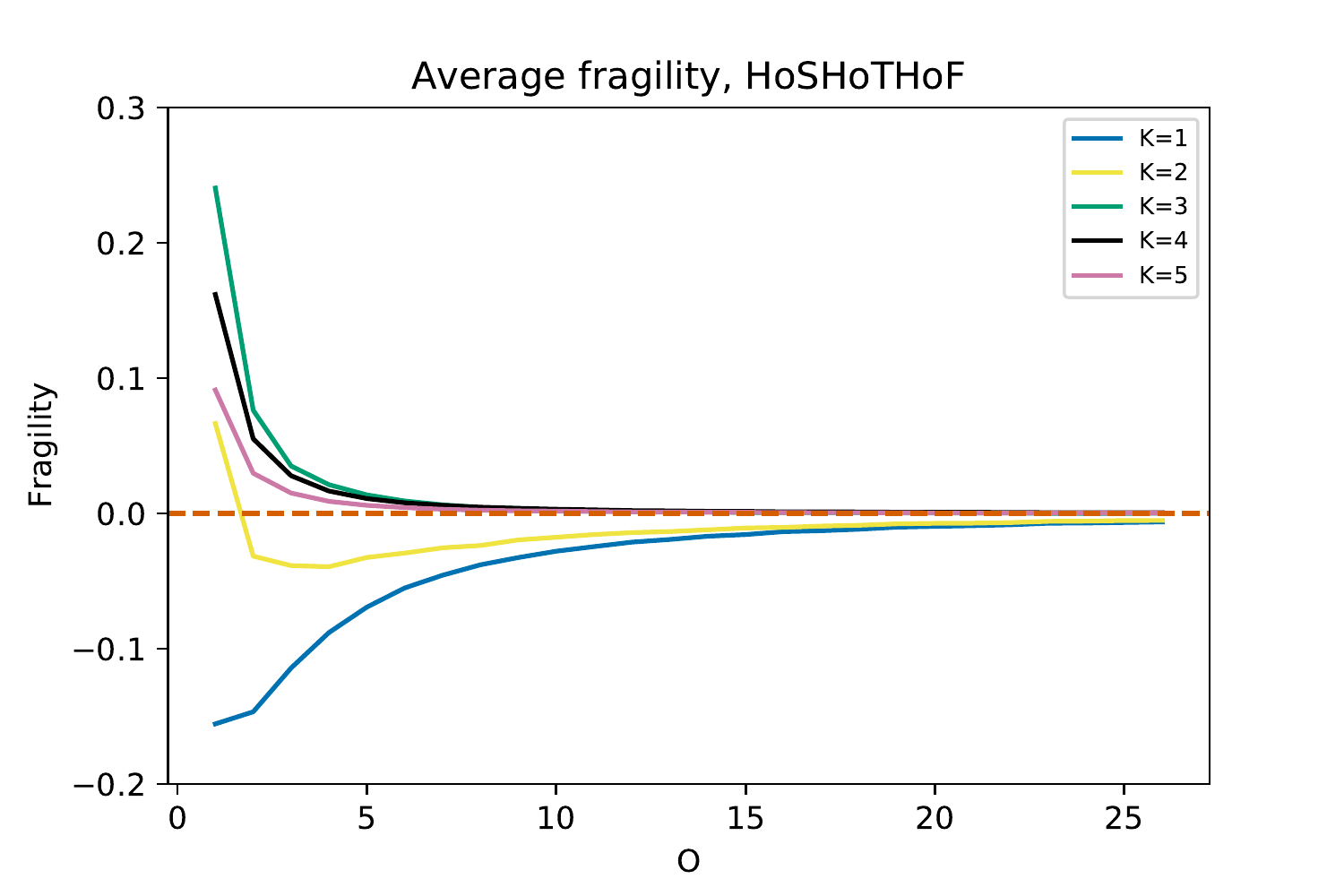} 
    \caption{ } 
    \label{AvsO_Ho} 
    \vspace{4ex}
  \end{subfigure} 
  \begin{subfigure}[b]{0.5\linewidth}
    \centering
    \includegraphics[scale=0.5]{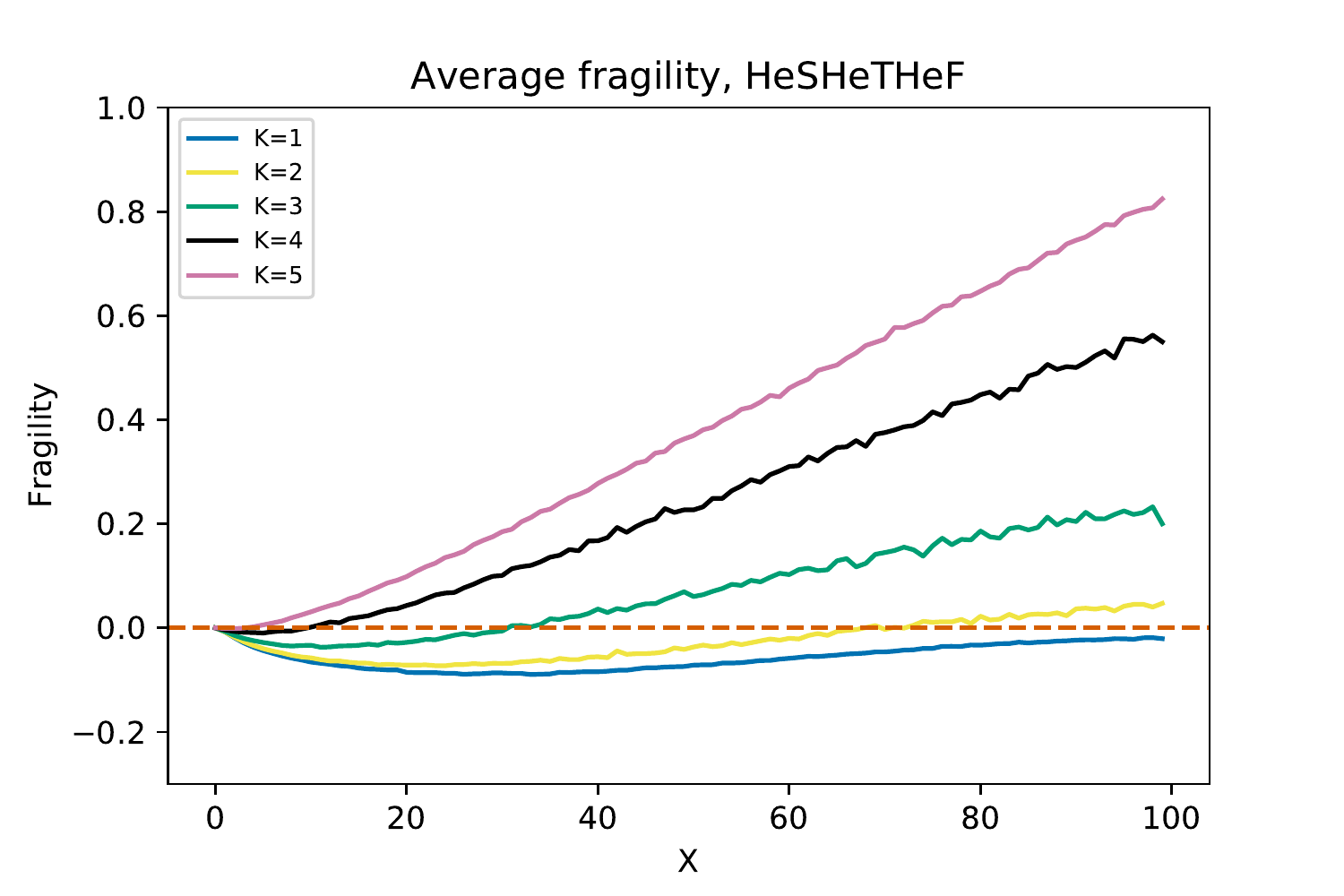} 
    \caption{ } 
    \label{AvsX_He} 
  \end{subfigure}
   \begin{subfigure}[b]{0.5\linewidth}
    \centering
    \includegraphics[scale=0.5]{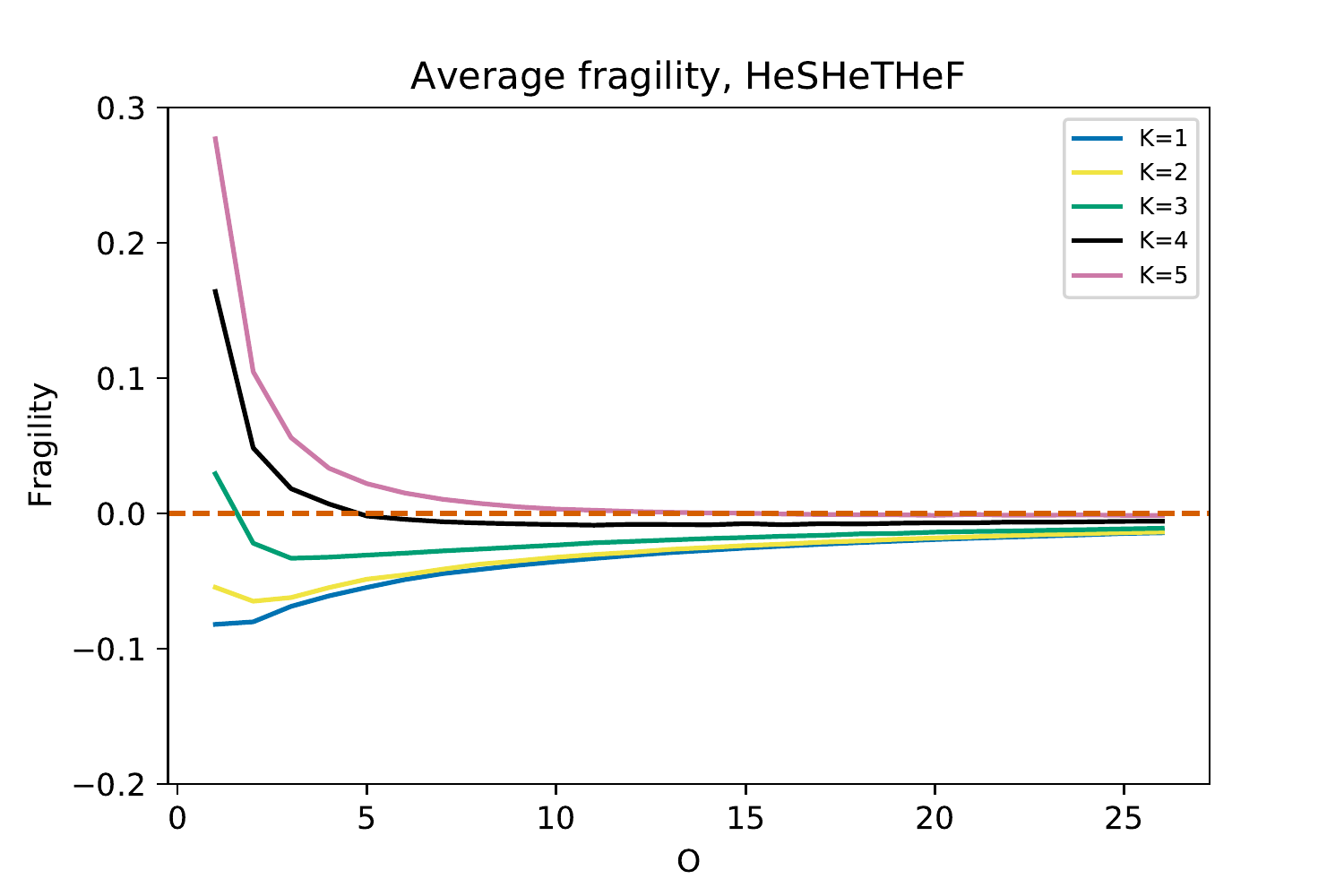} 
    \caption{ } 
    \label{AvsO_He} 
  \end{subfigure}
  \caption{Average antifragility of ordered, critical, and chaotic RBNs depending on $X$ and $O$ for 3Ho and 3He cases. In the homogeneous case (\ref{AvsX_Ho},~\ref{AvsO_Ho}) only the ordered and critical networks achieve antifragility intervals, the former reaching much larger values. Adding triple heterogeneity (\ref{AvsX_He},~\ref{AvsO_He}) shows a reversal in the order of the previously chaotic networks, and in this case two of the three curves for $K>2$ reach antifragility intervals. Although in the heterogeneous case the number of curves attaining antifragility increases and the length of the intervals with antifragility extends, in the homogeneous case much larger antifragility values are obtained. For all curves, steps of $\Delta X = 1$, $\Delta O = 1$ were used and in every step a total of 1000 iterations were averaged.}
  \label{Anti}
\end{figure}

So far we have explored how heterogeneity extends criticality, which had been partially studied in~\cite{Sanchez-Puig2022}. Another property that is fundamental to the evolution and adaptability of complex systems is antifragility, so it is worth studying whether heterogeneity also extends antifragility in some way. Figure~\ref{Anti} shows the variation of fragility with respect to $X$ and $O$ for five different connectivity values. For the 3Ho case, $K=1$ represents ordered dynamics, $K=2$ represents critical dynamics and $K=3, 4, 5$ represents chaotic dynamics. When the curves become negative this implies  antifragility. In the homogeneous case (Fig.~\ref{AvsX_Ho} and~\ref{AvsO_Ho}) both the ordered network and the critical network show increased antifragility, which is absent in the chaotic cases. By adding a triple heterogeneity (Fig.~\ref{AvsX_He} and~\ref{AvsO_He}) we observe an extension in the antifragility intervals for Fragility vs $O$ and for Fragility vs $X$. Despite the extension of the antifragility intervals, we note that for the homogeneous case, larger values of antifragility are reached, which again indicates a necessary balance between homogeneity and heterogeneity. On the other hand, with 3He for $K>2$ other interesting things happen. First, we observe that the curves for $K=4$ remain almost invariant to the change from homogeneity to heterogeneity. Moreover, there is a reversal of order in the curves for $K=3$ and $K=5$, since for the homogeneous case $K=3$ is always the upper curve, but for the heterogeneous case $K=5$ becomes predominant and even appear regimes where $K=3$ and $K=4$ reach antifragility. These results are consistent with those found in~\cite{Pineda2019}. Heterogeneity not only extends criticality, but also the region of ordered dynamics, where antifragility is highest. However, homogeneous networks are ``more ordered'' than heterogeneous ones. Thus, analogously with criticality, a balance between homogeneity and heterogeneity is also necessary to achieve ``optimal'' antifragility in complex systems.

\section{Conclusions and Future Work}

An open cosmological problem is the existence of a fine-tuned universe~\cite{monton2006god}, where only certain values in the fundamental constants give rise to the complex structures that surround and sustain us. This has evoked a series of different explanations. One approach proposes that heterogeneity offers a natural way to extend criticality. As such, this perspective is consistent with the fact that diverse natural structures and social architectures are heterogeneous. Our work shows that by adding some kind of heterogeneity to RBNs (structural, temporal, or functional) both criticality and antifragility are extended. The generality of this model suggests that the results obtained could be extended to different natural and social phenomena that can be modeled with networks. Although heterogeneity may not be the only way to increase the criticality or antifragility of a system, the results obtained show that once heterogeneity is present, the dynamics will have certain evolutionary advantages. This implies that parameter fine-tuning is not necessary to obtain the benefits of criticality, making the evolution of complex systems faster and more reliable. It is also important to emphasize that despite the benefits obtained, heterogeneity is not ``the whole enchilada''. Simulations show that a balance between homogeneity and heterogeneity is necessary for a complex system to thrive in a dynamic environment.

In our explorations there are many parameters that could be varied. Certainly, we could perform a more thorough investigation of the parameter space, but there are also other interesting things to explore. The results point to the existence of a hierarchy in the heterogeneities. Each heterogeneity has a different weigh for extending criticality and antifragility. We do not have an explanation for this hierarchy. Why is HeF less effective at extending criticality than HeS? Why does HeS produce greater complexity than HeT even though HeT seems to give more structure to node transitions? Implementing other types of heterogeneity (and their combinations) could provide answers to these questions. Could it be that the combinations between homogeneities and heterogeneities induce a hierarchy? What would be the rules governing this ranking?~\cite{Iniguez2021}

When varying the standard deviations in the Gaussian distribution, a clear tendency of the curves to converge after a certain value of $\sigma$ was noticed. Although this behavior does not seem universal, this same phenomenon could be studied in other functional distributions. Likewise, when varying the mean we found a symmetry with respect to $\mu=0.5$ and in fact we could also investigate if this symmetry is present in other distributions and in non-Boolean networks~\cite{SoleEtAl2000}. If so, we could argue that there is a general symmetry and, as any continuous symmetry, this induces a conservation law (which can be highly non-trivial~\cite{kosmann2011noether}). Then there would be a general conserved quantity that governs the dynamics of the system and that can help us to better understand how complex systems work in general. Also, the parameter that we investigated in the less furtive way was the temporal updating. Perhaps other temporal updating schemes (including some variations of the Out-degree mechanism) could be devised to find out which is the best strategy in the temporal heterogeneous case.

So far we have shown that, although the presence of heterogeneity is important in extending criticality and antifragility, this rule is not entirely universal. Under certain conditions, decreasing heterogeneity and increasing homogeneity can also be beneficial in maximizing both criticality and antifragility. This is even more visible when adding functional heterogeneity, since in this case it can be argued that heterogeneity is not entirely additive, i.e., it is not true that the more heterogeneity the higher the criticality threshold. Also, we can appreciate a relationship between the number of nodes in the network ($N$) and the connectivity among them ($K$). For networks with few nodes, it takes a larger $K$ to take advantage of this heterogeneity overlap. At the same time, if $N$ is large enough, the connectivity from which we take advantage in the 3He case decreases. During the evolution of any complex system there may be phases where conditions are not optimal for exploiting environmental resources, which reinforces the idea of balance described above. The functional and temporal variations reflect the high dependence between the variables chosen and the complexity curves obtained. For now, it seems difficult to state that a certain combination of parameters produces maximum criticality and this is mostly due to the computational expense involved in working with  large $K$ values. 

Despite their generality, it is not possible to model all natural and social phenomena using random Boolean networks. One of the major limitations of these models is that they only allow each node to have two states. However, it is not difficult to imagine dynamics in which the binary nature of the states is absent. Implementing networks with more than two states is a computational challenge, since even in the Boolean case the memory can be filled exponentially. Nevertheless, there are ecological models such as Roy et al.~\cite{roy2003broad} that show the importance of heterogeneity for two-dimensional lattices whose nodes can have three different states. With sufficient computational power, it will be possible to explore the effects of heterogeneity in much more general networks with non-binary states~\cite{Zapata2014Random-Fuzzy-Ne,Zapata2020}. Another difficulty in sketching nature is the entanglement between its different scales. In general, networks only allow to capture the system at one of its scales. If one wishes to study a phenomenon at different scales (and the interactions between them) using networks, concepts such as causal emergence have been developed~\cite{hoel2013quantifying}. It is also possible to attack this problem by relating spatial structure and global densities, an example of this that also reflects the importance of heterogeneity is studied in ecology by Pascual et al.~\cite{pascual2011simple}. Finally, in many network models the connectivity between nodes is fixed. This implies that elements can only interact with a fixed set of components throughout the entire dynamics. This dynamic rigidity can be abolished by using, for example, agent-based models, and more generally with adaptive networks~\cite{gross2009adaptiveNets}. Although criticality and resilience have already been studied in these models~\cite{khajehabdollahi2022critical}, the effects of heterogeneity on them remain to be studied.


\vspace{6pt} 



\authorcontributions{All authors conceived and designed the study. F.S.P. wrote the software. A.J.L.D. 
performed numerical simulations and derived mathematical results. All authors wrote the paper.}

\funding{C.G. acknowledges support from UNAM-PAPIIT (IN107919, IV100120, IN105122) and from the PASPA program from UNAM-DGAPA.}

\institutionalreview{Not applicable.}

\acknowledgments{We appreciate useful comments from Dante Chialvo, Gerardo I\~niguez, and G\'eza \'Odor.}

\conflictsofinterest{The authors declare no conflict of interest.}


\sampleavailability{Samples of the compounds ... are available from the authors.}


\abbreviations{Abbreviations}{
The following abbreviations are used in this manuscript:\\

\noindent 
\begin{tabular}{@{}ll}
MDPI & Multidisciplinary Digital Publishing Institute\\
DOAJ & Directory of open access journals\\
RBN & Random Boolean Network
\end{tabular}
}




\begin{adjustwidth}{-\extralength}{0cm}

\reftitle{References}


\bibliography{Definitions/amahury}

\end{adjustwidth}
\end{document}